\documentclass[a4paper,fleqn]{cas-dc}

\usepackage[numbers]{natbib}
\usepackage{threeparttable}
\usepackage{multirow}
\usepackage{makecell}
\usepackage{caption}
\usepackage{subcaption}
\usepackage[figuresright]{rotating}
\usepackage{amssymb}
\usepackage{graphicx}
\usepackage{subcaption}
\usepackage{rotating} 
\usepackage{tabularx} 
\usepackage{lscape} 
\usepackage{amsmath}
\usepackage{algorithm}
\usepackage{algorithmicx}
\usepackage{algpseudocode}
\usepackage{epsfig}
\def\tsc#1{\csdef{#1}{\textsc{\lowercase{#1}}\xspace}}
\tsc{WGM}
\tsc{QE}
\begin{document}
%%%%封面内容编辑%%%%
\begin{titlepage} % Suppresses headers and footers on the title page

	\centering % Centre everything on the title page
	
	\scshape % Use small caps for all text on the title page
	
	\vspace*{\baselineskip} % White space at the top of the page
	
	%------------------------------------------------
	%	Title
	%------------------------------------------------
	
	\rule{\textwidth}{1.6pt}\vspace*{-\baselineskip}\vspace*{2pt} % Thick horizontal rule
	\rule{\textwidth}{0.4pt} % Thin horizontal rule
	
	\vspace{0.75\baselineskip} % Whitespace above the title
	
	{\LARGE Could Bibliometrics Reveal Top Science and Technology Achievements and Researchers? The Case for Evaluatology-based Science and Technology Evaluation.} % Title
	
	\vspace{0.75\baselineskip} % Whitespace below the title
	
	\rule{\textwidth}{0.4pt}\vspace*{-\baselineskip}\vspace{3.2pt} % Thin horizontal rule
	\rule{\textwidth}{1.6pt} % Thick horizontal rule
	
	\vspace{2\baselineskip} % Whitespace after the title block
	
	%------------------------------------------------
	%	Subtitle
	%------------------------------------------------
	
	%Subtitle here % Subtitle or further description
	
	\vspace*{3\baselineskip} % Whitespace under the subtitle
	
	%------------------------------------------------
	%	Editor(s)
	%------------------------------------------------
	
	Edited By
	
	\vspace{0.5\baselineskip} % Whitespace before the editors
	
	{\scshape\Large Guoxin Kang\\ Wanling Gao\\ Lei Wang\\ Chunjie Luo\\Hainan Ye\\Qian He\\Shaopeng Dai\\Jianfeng Zhan\\}

	\vspace{0.5\baselineskip} % Whitespace below the editor list

	\vfill % Whitespace between editor names and publisher logo
	
	%------------------------------------------------
	%	Publisher
	%------------------------------------------------
	
	%\plogo % Publisher logo
	%\def\BUlogo{\epsfig{file=ICT.pdf,height=3cm}}
	%\includegraphics[scale=0.135]{ICT.pdf}
	\epsfig{file=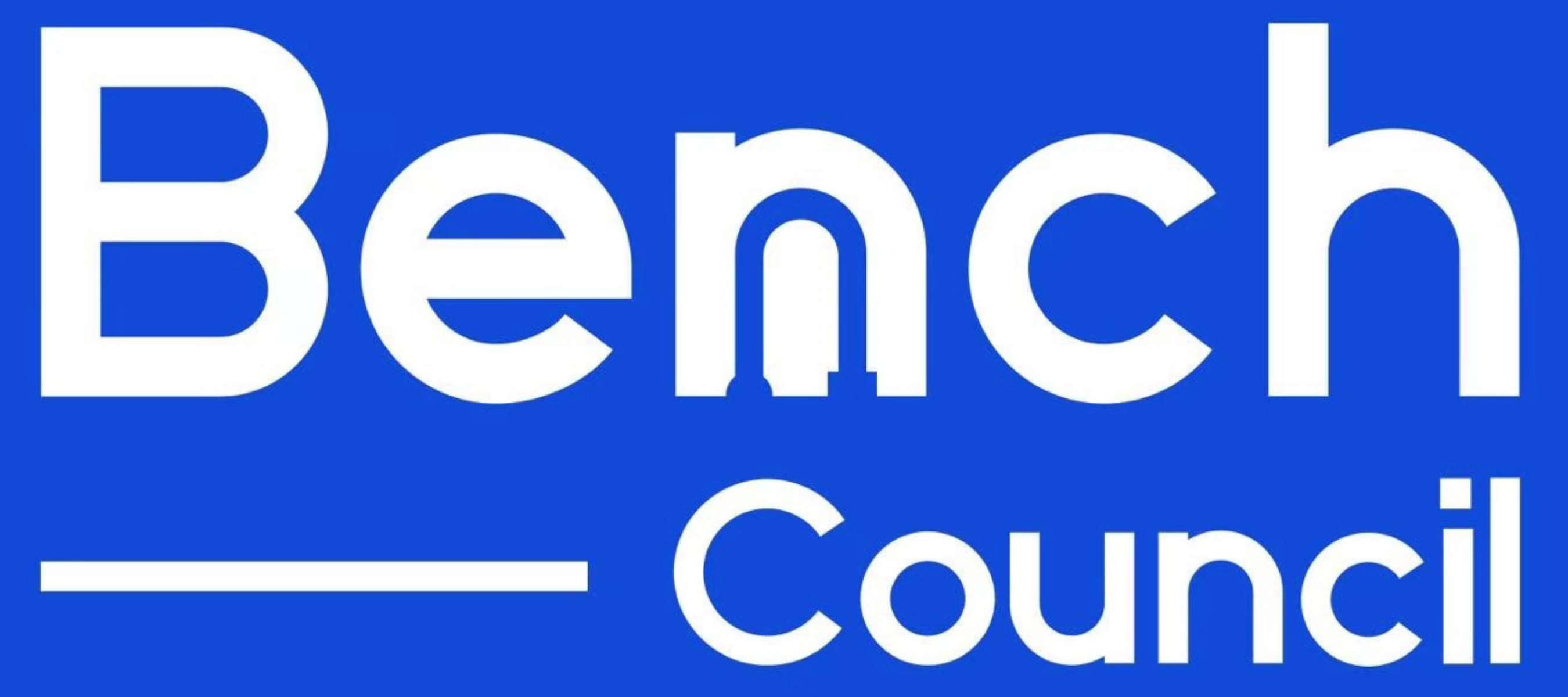,height=2cm}
	\textit{\\BenchCouncil: International Open Benchmark Council\\http://www.benchcouncil.org} % Editor affiliation
	\vspace{5\baselineskip} % Whitespace under the publisher logo

	Technical Report No. BenchCouncil-S\&T Evaluatology-2024 % Publication year
	
	{\large Aug 21, 2024} % Publisher

\end{titlepage}

%----------------------------------------------------------------------------------------

\let\WriteBookmarks\relax
\def\floatpagepagefraction{1}
\def\textpagefraction{.001}

% Short title
\shorttitle{Could Bibliometrics Reveal Top Science and Technology Achievements and Researchers? The Case for Evaluatology-based Science and Technology Evaluation. }    

% Short author
\shortauthors{Dr. Guoxin Kang et al}

% Main title of the paper
% \title [mode = title]{Inference Accuracy Variability}
\title [mode = title]{Could Bibliometrics Reveal Top Science and Technology Achievements and Researchers? The Case for Evaluatology-based Science and Technology Evaluation. }

\author[2,1,3]{Guoxin Kang}
\credit{Contributions to the summary of the related work for CSRankings and c-score in Section 2.2, the whole-session discussion, the mathematical formulations of four relationships, the algorithm for identify the four fundamental relationships in Section 3.3, the benchmark examples in Section 3.3, and the  presentations of Figure 1, 2, 3 and Table 1}
\author[2,1,3]{Wanling Gao}
\credit{Contributions to the summary of the related work for H-index in Section 2.2, the whole-session discussion, partial revision of mathematical formulations and algorithms in Section 3.3, and the presentations of Figure 4, 5, 6, 7}
\author[2,1,3]{Lei Wang}
\credit{Contributions to the summary of the related work for CiteScore and SNIP in Section 2.2, the whole-session discussion, the writing of Section 4, Section 5, the chip examples in 3.3, and the presentations of Figure 5, 6}
\author[2,1,3]{Chunjie Luo}
\credit{Contributions to the part of SJR in related work, the AI examples in Section 3.3, and the discussion}
\author[1,2,3]{Hainan Ye}
\credit{Contributions to the data for Chip100 rankings and the discussion}
\author[1]{Qian He}
\credit{Contributions to the presentations of Figure 8 and the discussion}
\author[3]{Shaopeng Dai}
\credit{Contributions to the discussion}
\author[1,2,3]{Jianfeng Zhan}
\credit{Contributions to the proposal for the evaluatology-based science and technology evaluation methodology, including the extended EC, four relationships, real-world ES,  perfect S\&T EM, pragmatic S\&T EM, and Top N @X @Y methodology. Contributions to the presentation of most texts, excluding figures and tables. Also contributions to other aspects of the work unless otherwise explicitly stated}
\ead{jianfengzhan.benchcouncil@gmail.com}
\ead[url]{www.zhanjianfeng.org}

\address[1]{The International Open Benchmark Council}
\address[2]{ICT, Chinese Academy of Sciences, Beijing, China}
\address[3]{University of Chinese Academy of Sciences, Beijing, China}

\begin{abstract}
By utilizing statistical methods to analyze bibliographic data, bibliometrics faces inherent limitations in identifying the most significant science and technology achievements and researchers. To overcome this challenge, we present an evaluatology-based science and technology evaluation methodology.
At the heart of this approach lies the concept of an extended evaluation condition, encompassing eight crucial components derived from a field.  We define four relationships that illustrate the connections among various achievements based on their mapped extended EC components, as well as their temporal and citation links. Within a relationship under an extended evaluation condition, evaluators can effectively compare these achievements by carefully addressing the influence of confounding variables.
% the specific problem domains within Implementing a well-defined extended evaluation condition on particular achievements constitute evaluation models or systems. 
We establish a real-world evaluation system encompassing an entire collection of achievements, each of which is mapped to several components of an extended EC. Within a specific field like chip technology or open source, we construct a perfect evaluation model that can accurately trace the evolution and development of all achievements in terms of four relationships based on the real-world evaluation system. Building upon the foundation of the perfect evaluation model, we put forth four-round rules to eliminate non-significant achievements by utilizing four relationships. This process allows us to establish a pragmatic evaluation model that effectively captures the essential achievements, serving as a curated collection of the top N achievements within a specific field during a specific timeframe.
We present a case study on the top 100 Chip achievements to demonstrate the effectiveness of our science and technology evaluatology. The case study highlights its practical application and efficacy in identifying significant achievements and researchers that otherwise can not be identified by using bibliometrics.
\end{abstract}

\begin{keywords}
 \sep Evaluatology
 \sep Science and technology evaluation
 \sep Top achievements and researchers
 \sep Bibliometrics
 \sep Extended evaluation condition  
 \sep Evaluation systems or models
 \sep Top N @X @Y  
\end{keywords}

\maketitle
\section{Introduction}

Science and technology (S\&T) evaluation is a meticulous and comprehensive process. One of its paramount goals is to identify the most remarkable accomplishments in each field, duly recognize the individuals, institutions, or nations that have made significant contributions to these achievements, and delve deeper into the effective and efficient mechanisms and policies within the S\&T ecosystems that profoundly shape the evolution of these achievements~\cite{evaluation}. This article focuses on the first half of the task.

%Secondly, it entails examining the impact of critical factors within the science and technology ecosystem on these outcomes, aiming to advance scientific knowledge and technological progress.

\begin{figure}
	\centering
\includegraphics[scale=.1]{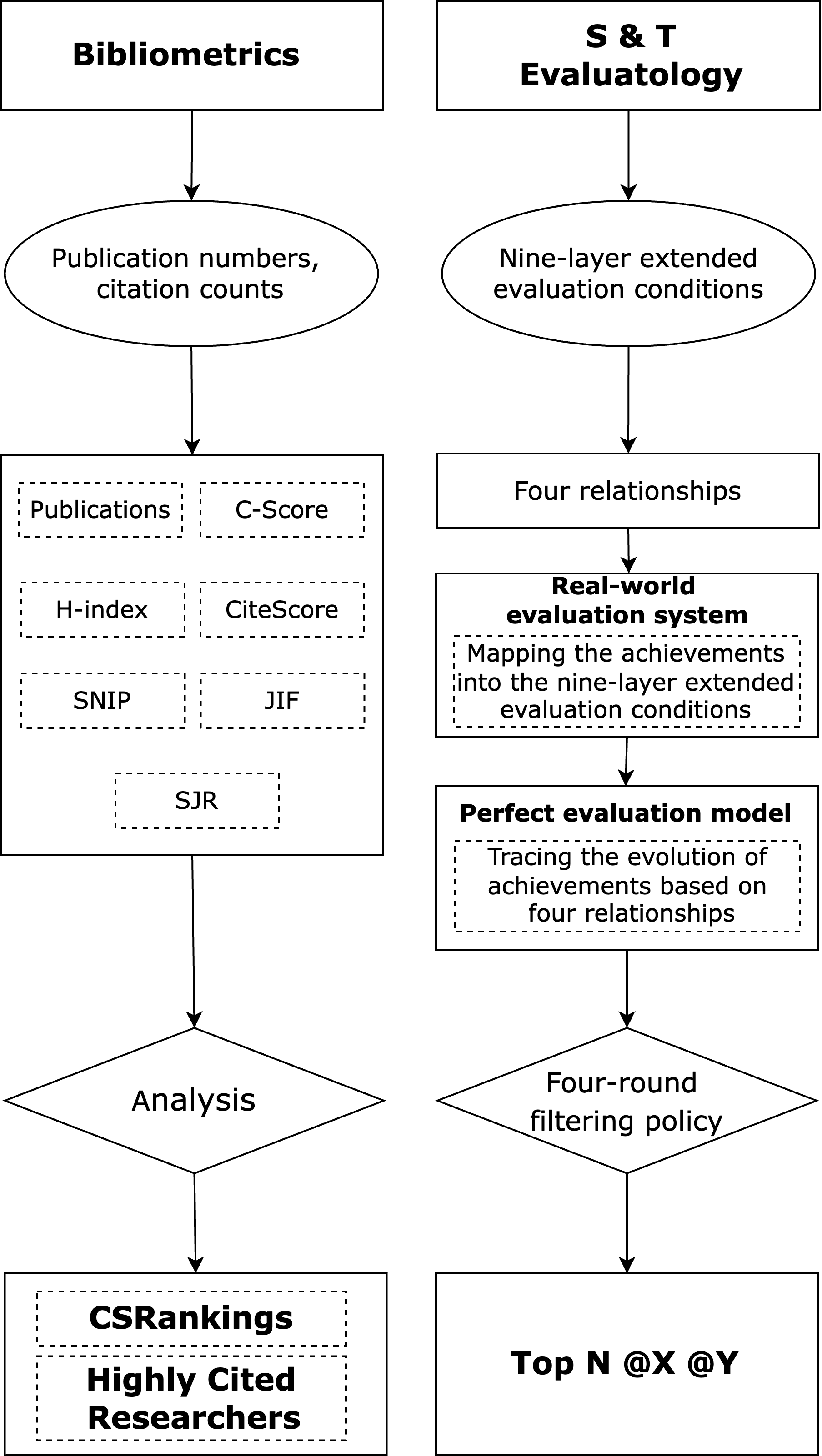}
	\caption{Fundamental differences between bibliometrics and S\&T Evaluatology.}
	\label{evaluatology}
\end{figure}

While bibliometrics methodologies have long relied on observable metrics such as publication numbers, citation counts, and the H-index to assess correlations and impact~\cite{CSRankings, ioannidis2019standardized, hirsch2005index, hindexwiki}, as illustrated in Figure ~\ref{evaluatology}. it is essential to recognize their inherent three limitations and the need for alternative approaches.

First, bibliometrics commonly employs publication numbers, citation counts, and related metrics to gauge scholarly works' quality, influence, and significance. However, various confounding variables can significantly impact citation counts.  Moreover, citation counts are vulnerable to manipulation by malicious networks.

Second,  bibliometrics often fails to consider critical non-bibliometric metrics, making them insufficient for evaluating significant technological achievements that may have limited publication outputs. For instance, the Linux operating system in computer science has made a substantial impact despite having a modest publication record.

Third, many bibliometrics methodologies prioritize the quantity over the quality of publications, which can result in an incomplete assessment of the true value and impact of scholarly work.

To address these shortcomings, we introduce the S\&T evaluatology, which exemplifies the application of evaluatology in evaluating S\&T achievements.
The S\&T evaluatology is illustrated in Figure ~\ref{evaluatology} and presented in detail in~\cite{Evaluatology, A_short_summary_Evaluatology}. The fundamental principle of evaluatology is to implement a well-defined evaluation condition (EC) on particular subjects to establish evaluation models or systems. 

%Evaluators can deduce the effects of these subjects under a well-defined EC by skillfully addressing the impact of confounding variables.  

At the core of the S\&T evaluatology is the notion of an extended EC, which comprises nine key components: (1) the field that can be broken down into several problem domains; (2) The set of problem domains, each of which can be broken down into various sub-problem domains; (3) the sub-problem domains, each of which can be decomposed into several problems; "(4) the set of a collective of equivalent problems, each of which can be broken down into multiple sub-problems; (5) the set of a collective of equivalent sub-problems;  (6) the set of a collective of problems or sub-problem instances; (7) the algorithms or the algorithm-like mechanisms that tackles a problem or a sub-problem; (8) the implementations of algorithms or the algorithm-like mechanisms; (9) the support systems that provide necessary resources and environments~\cite{A_short_summary_Evaluatology, Evaluatology}". 

We define four relationships that illustrate the connections among various achievements based on their mapped extended EC components, as well as their temporal and citation links. 
We define two primary relationships: ~\textit{pioneering and progressive} and two auxiliary relationships: ~\textit{parallel} and ~\textit{related but not connected}.
 Within a pioneering or progressive relationship under an extended evaluation condition, evaluators can effectively compare these achievements by carefully addressing the influence of confounding variables.

We establish a real-world ES encompassing the complete collection of S\&T achievements,  each of which is mapped to several components of an extended EC.  In line with the aim of identifying the top N S\&T achievements, the proposed real-world S\&T ES ignores the other components of the real-world S\&T ecosystems, e.g., the mechanisms and policies that profoundly shape the evolution of these achievements~\cite{evaluation}.

Under the premise that all evaluated achievements belong to the same field, e.g., chip technology or open source, we construct "a perfect S\&T EM" that can accurately trace the evolution and development of all achievements in terms of four relationships based on the real-world ES. We compare achievements that have a specific relationship under the extended EC they involve. Utilizing four relationships, we employ four rounds of rules to prune non-significant achievements to establish a pragmatic S\&T EM that captures the fundamental S\&T achievements. Essentially, the pragmatic S\&T EM is a collection of top N achievements within a field during a timeframe. 

The International Open Benchmark Council (BenchCouncil) utilized the S\&T evaluatology principles and the instantiated Top N @X @Y methodology to systematically recognize the most 100 groundbreaking and influential achievements in chip technology (Chip100)~\cite{Chip100}. The case study demonstrates the effectiveness of our proposed methodology compared to bibliometrics.

In the following sections, we will provide an in-depth examination of the S\&T evaluatology. Section~\ref{related} enumerates the existing bibliometrics methodologies and analyzes their weakness. Section~\ref{metric} presents the S\&T evaluatology in detail. Section~\ref{Top_N_X_Y} provides an instantiated Top N @X @Y methodology. Section~\ref{ranking} introduces a case study on the Top 100 Chip Achievements. 
%Section~\ref{discussion} discusses why our proposed methodology overcomes the limitations of bibliometrics. 
Section~\ref{conclusion} concludes.
\section{Motivation and Related Work}~\label{related}

Bibliometrics is a field that applies statistical methods to analyze bibliographic data. 

In this subsection, we first present the overall weakness of bibliometrics in Section~\ref{bibliometrics_weakness}. Then, we introduce the representative bibliometrics methodologies. Finally, we introduce the fundamental concept, theory, and methodology in evaluatology~\cite{Evaluatology}, based on which we will present the S\&T evaluatology. 

\subsection{Motivation: The limitations of bibliometrics}~\label{bibliometrics_weakness}

Due to the nature of bibliometrics, there are several inherent drawbacks associated with its application. 

      \begin{figure}
	\centering
		\includegraphics[scale=.57]{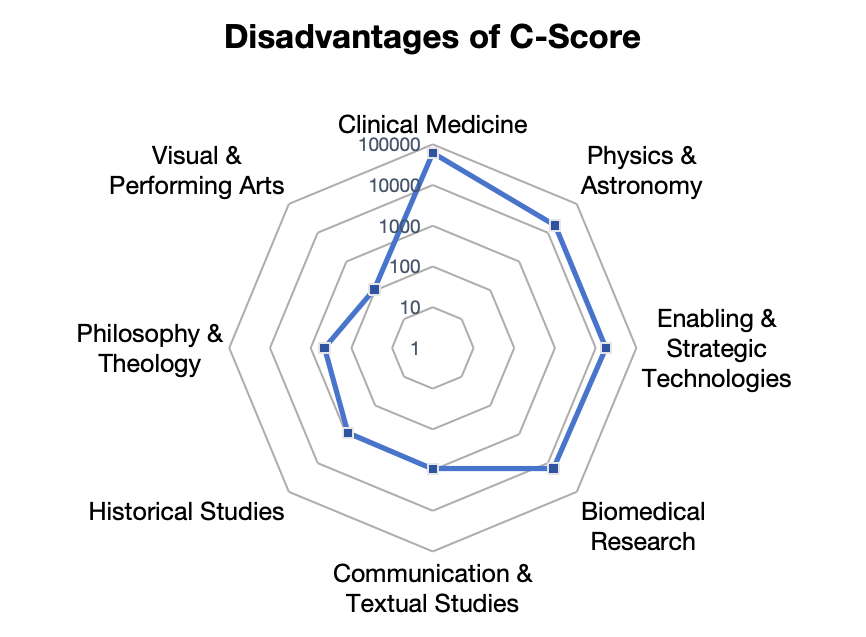}
	\caption{Comparison of the number of scientists selected for the global top 2\% in different disciplines.}
	\label{cscore}
\end{figure}

First,  bibliometrics commonly employs publication numbers, citation counts, and related metrics to gauge the quality, influence, and significance of scholarly works. However, it is crucial to acknowledge that publication numbers and citation counts can be significantly impacted by various confounding variables. These may include the diverse disciplines involved, the reputation and networks linked to researchers and their institutional affiliations, as well as notable differences in researcher numbers and publication volumes across various fields. Moreover, citation counts may even be vulnerable to manipulation by malicious networks (Limitation One).

\begin{itemize}

 \item  (Limitation One-One). The same or similar works published in the same or different periods can be impacted by various confounding variables, such as the reputation and network of the researchers and their institutions, leading to significant variations in citation counts. Moreover, citation counts may be subject to manipulation by malicious networks. 

    \item (Limitation One-Two). In fundamental disciplines like mathematics, once a problem has been effectively solved, there may be limited follow-up research on that specific topic. Consequently, the citation count for the original work in fundamental disciplines may not increase significantly. Hence, it can not accurately reflect the impact or influence of the research in the field.

      \item (Limitation One-Three). Citation counts fail to account for the significant disparities in researcher numbers and publication volumes across different fields. In fields with fewer researchers and publications, citation counts are naturally lower, regardless of the quality of the research being conducted. 
    
      \item (Limitation One-Four). Bibliometrics prioritize well-established disciplines, potentially overlooking emerging fields or unconventional research outputs that may have a significant impact but lower citation counts.
            The effectiveness of citation metrics is limited in representing contributions within emerging or specialized fields. Groundbreaking research in these domains might initially receive few citations due to the novelty of the subject matter or the field's limited scope. Consequently, as shown in Figure~\ref{cscore}, pivotal advancements in such areas risk being undervalued, as seen in the "top scientists" list created by Stanford University and the Elsevier data repository~\cite{ioannidis2019standardized}, which predominantly features scientists from well-established fields like Clinical Medicine and Physics \& Astronomy. This bias is particularly harmful to innovators who spearhead new research directions, as their contributions may not be accurately captured by citation-based metrics.

   \item  (Limitation One-Five). Self-citations occur when authors cite their previous work, potentially inflating the impact of their research.  This practice skews the representation of a paper’s or a researcher’s genuine influence within the academic community. For instance, metrics like the H-index~\cite{hirsch2005index} are unable to circumvent the issue of self-citations, resulting in a biased assessment that may unfairly favor those who self-cite frequently.

\end{itemize}

Second, bibliometrics totally ignores other fundamental non-bibliometric metrics and hence can not be applied to significant technological achievements that have few or no publication outputs (Limitation Two). Bibliometrics primarily relies on analyzing published works. Limitation Two arises when considering groundbreaking technological advancements that may not be adequately represented in traditional scholarly publications. 
 
 In practical fields like computer science, substantial contributions frequently occur outside the conventional academic publishing framework.  A prime example of this is the Linux operating system within the realm of computer science. As an open-source software, the Linux operating system boasts numerous contributors who may not publish extensively.
  Similarly, the computer mouse, one of the most universally adopted human-computer interaction technologies, demonstrates that significant impact does not necessarily stem from published research.  Table~\ref{non_bibiolimetric_example} presents several significant technological achievements that are overlooked by bibliometrics. 
Therefore, bibliometrics alone may not fully capture the impact and significance of these achievements. Consequently, non-academic metrics should be considered in the grading process to select the top-impact achievements. 

Third, bibliometrics prioritizes the quantity over the quality of publications (Limitation Three).  High citation counts of a researcher can result from either a large volume of modestly impactful publications or from many surveys on trending topics, such as timely topics on large language models.  Although these works might garner significant attention, they do not necessarily represent substantial advancements within their disciplines. The emphasis on publication counts can lead to a skewed representation of research impact, as it fails to consider the significance, rigor, and originality of individual publications.

In summary, while bibliometrics provides a quantitative metric,  like citation counts, for academic evaluation, they are beset with limitations that result in biased and incomplete assessments.
Thus, the S\&T evaluation urgently needs more nuanced and comprehensive evaluation metrics and methodologies that go beyond bibliometrics. Such metrics would ensure a fairer and more accurate depiction of scholarly impact, truly reflecting the multifaceted nature of academic contributions. 

\begin{table}[]
\caption{Summary of significant achievements overlooked by bibliometrics}
\begin{tabular}{|p{1.7cm}|p{1.7cm}|p{1.2cm}|p{1.2cm}|}
\hline
Field                                                                           & Achivements & Published Paper & Citations \\ \hline
\multirow{2}{*}{Chip}                                                           & X86 ISA   & No              & N/A            \\ \cline{2-4} 
                                                                                 & PCB       & No              & N/A            \\ \hline
%AI                                                          %                     &             &                 &      %          \\ \hline
\multirow{3}{*}{\begin{tabular}[c]{@{}c@{}}Open-sources \\ systems\end{tabular}} & Linux Kernel      & No              & N/A            \\ \cline{2-4} 
                                                                                 & Git         & No              & N/A            \\ \cline{2-4} 
                                                                                 & MySQL       & No              & N/A            \\ \hline
\multirow{4}{*}{Benchmarks}                                                      & Whetstone   & No              & N/A            \\ \cline{2-4} 
                                                                                 & TPC-C       & No              & N/A            \\ \cline{2-4} 
                                                                                 & TPC-H       & No              & N/A            \\ \cline{2-4} 
                                                                                 & FIO         & No              & N/A            \\ \hline
\end{tabular}
\label{non_bibiolimetric_example}
\end{table}
\subsection{The representative bibliometrics methodologies}

\subsubsection{CSRankings in the computer science field}
~\label{related-csranking}

CSRankings is a specialized method for evaluating computer science achievements, favoring the conference publication over the journal. 
CSRankings adopts the metric of the number of publications at so-called top-tier conferences for gauging the academic influence of researchers or their affiliated institutions in computer science. Utilizing this metric, Emery Berger pioneered CSRankings~\cite{CSRankings}, a tailored academic leaderboard specifically designed for the realm of computer science. CSRankings selects the Digital Bibliography \& Library Project (DBLP)~\cite{DBLP} as its data source, ensuring up-to-date and relevant rankings with quarterly updates.

 However,  this methodology has several serious flaws.  
First and foremost, it places a higher emphasis on publication quantity than quality, as outlined in Section~\ref{bibliometrics_weakness} (Limitation Three), thereby having flaws in recognizing top researchers or groundbreaking achievements.
 
 For example, David Patterson's influential works in chip technology, particularly with RISC, RISC-V, and RAID, have substantially shaped the field. Notwithstanding their extensive influence, Patterson is conspicuously absent from CSRankings, a glaring omission highlighting a significant shortcoming in the ranking system's ability to acknowledge key contributors even in leading institutions.  

In addition,  CSRankings can not discern the varying impacts of different achievements.  CSRankings quantifies the number of papers presented at top-tier conferences, but this approach fails to identify who pioneered a field. For instance, although the groundbreaking ``Transformer`` model~\cite{vaswani2017attention} was presented at the 31st Conference on Neural Information Processing Systems (NeurIPS), it is erroneous to assume that all papers at this conference exert an influence comparable to that of the Transformer. 

This situation underscores a fundamental flaw in the CSRankings system: overemphasizing top-tier conference publications can lead to misleading representations, bypassing the real depth and enduring impact of substantial contributions. 

%Therefore, shifting from a narrow emphasis on conference publications to a more multi-dimensional metric for assessing achievement impact is essential. 
 
 Second, many influential works like the Linux operating system have never even sought publication in a so-called top-tier conference (Limitation Two). Table~\ref{non_bibiolimetric_example} provides other examples. The current metric focusing on publications fails to recognize significant achievements that are not encapsulated in conference papers. The Linux operating system's development and its widespread adoption stand as a prime example, achieving monumental impact without the endorsement of traditional academic publications. 

\begin{figure}
	\centering
		\includegraphics[scale=.57]{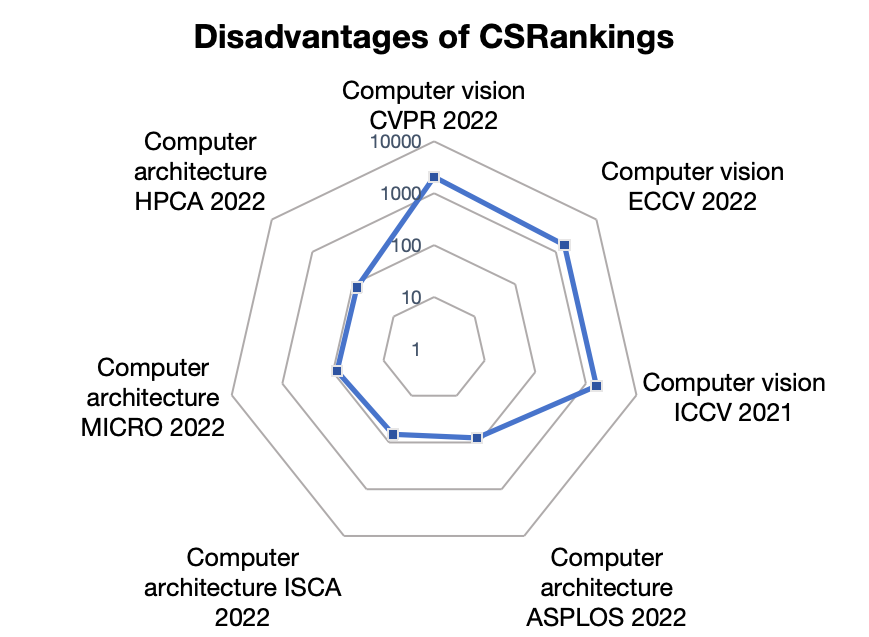}
	\caption{Comparison of accepted papers by top conferences in the fields of computer vision and computer architecture.}
	\label{csrangking}
\end{figure}

Third, CSRankings overlooks the significant disparities in researcher numbers, publication frequency, and volumes across different fields within computer science (Limitation One-Three). This oversight has resulted in a skewed ranking from 1970-2022, where four out of the top seven institutions are led by faculty specializing in vision, a field known for its high paper acceptance volumes. For example, the field of computer vision, known for higher publication volumes, is overrepresented. As shown in Figure~\ref{csrangking}, in 2022, The IEEE / CVF Computer Vision and Pattern Recognition Conference (CVPR), a leading conference in computer vision, accepted 2065 papers~\cite{cvpr2022}, whereas The IEEE/ACM International Symposium on Microarchitecture Conference (MICRO), a top conference in computer architecture, accepted just 83~\cite{micro2022}. 

The discrepancies in publication volume across different fields may lead to potential misleading outcomes when using CSRankings. These disparities raise concerns about the accuracy of CSRankings in providing equitable representation for all fields within computer science.

Fourth, as a consequence of being accepted by a so-called top-tier conference, this metric is impacted by various confounding variables, such as the reputation and network of the researchers and their institutions, and is subject to manipulation by malicious networks. Collusion among reviewers is not an isolated incident in numerous computer science conferences.

\subsubsection{The standardized citation metrics (c-score)}

The c-score, developed by John Ioannidis~\cite{ioannidis2019standardized}, assesses the influence of scientists. This standardized indicator amalgamates various elements, including citations, h-index, co-authorship-adjusted hm-index, and authorship-position-specific citations. Leveraging this metric, Ioannidis's team curated a global database for ranking scientists, categorized into career-long and single-year impacts based on the Scopus data. The former category spans citations from 1996 to now, while the latter focuses on the current calendar year alone. This innovative metric transcends traditional citation metrics, avoiding the evaluation biases introduced by self-citations. However, its primary focus on publications can not completely encompass the wider spectrum of a scientist's influence, particularly in areas such as practical applications or cross-disciplinary collaborations. These critical dimensions, essential to the fabric of scientific progress, are often understated in conventional bibliometric measures (Limitation Two).

Despite its popularity in the scientific field, the standardized citation metric has limitations in acknowledging the impact of researchers in emerging disciplines (Limitation One-Four), leading to an underrepresentation of their contributions.
The metric's proclivity to privilege well-established, voluminous fields is evidenced by the fact that over half of the top-ranked influential scientists in 2021 originated from fields like Clinical Medicine, Physics \& Astronomy, Biomedical Research, and Enabling \& Strategic Technologies. This trend reveals an inherent bias, favoring areas with more substantial publication frequencies and higher citation volumes (Limitation One-Three).

In addition, Limitation One remains a challenge that cannot be mitigated by the standardized citation metrics (c-score). Factors such as the reputation and network of researchers and their affiliated institutions can confound the evaluation process. For instance, even when two researchers from different institutions achieve similar achievements, the level of attention and recognition their work receives can vary significantly. In some cases, the earlier work may receive limited attention, while subsequent work gains widespread acclaim. These disparities can be attributed to various factors, including the visibility and influence that researchers and their institutions hold within the academic community.

\subsubsection{H-index} 

H-index~\cite{hirsch2005index} is a useful metric proposed by Jorge E. Hirsch to characterize a researcher's scientific output.  
The objective is to determine the highest value of h, where there are at least h papers with a citation number equal to or greater than h. The mathematical representation of H-index for a scientist is ${h\_index(f) = \displaystyle \max\{i\in \mathbb {N} :f(i)\geq i\}}$~\cite{hindexwiki}. Here, f is an array that contains the number of citations for the scientist's publications in decreasing order~\cite{hindexwiki}.
Instead of relying solely on single-number criteria like the total number of papers, H-index takes a more holistic approach by considering both productivity and academic impact. In addition to the limitations that we have discussed extensively in Section~\ref{bibliometrics_weakness}, in practice, vast self-citations can raise the H-index value easily.

%On the other hand, it is specific to academic evaluation and is primarily based on research paper publications.

\subsubsection{CiteScore metrics}

CiteScore Metrics~\cite{james2018citescore}, developed by Elsevier, extensively evaluates academic journals' citation impact and influence. These metrics are calculated yearly, considering a three-year citation window and considering the volume, quality, and field-normalized citation rates of articles published in a specific journal. Featuring indicators such as average citations per document, quartile ranking, and overall standing, CiteScore Metrics provides a transparent and comprehensive tool for researchers and institutions. While CiteScore metrics are designed to assess the quality and impact of scholarly journals rather than evaluate the quality of research within specific fields. 
In addition to the limitations that we have discussed in Section~\ref{bibliometrics_weakness}, it has another serious limitation. It is based on a three-year citation window. Consequently, achievements with a substantial long-term impact but relatively few citations in the short term may be undervalued.
%Moreover, Valenzuela et al.~\cite{valenzuela2015identifying} utilized a supervised classification method to identify the highly influential citations.

\subsubsection{Source Normalized Impact per Paper (SNIP)}

The Source Normalized Impact per Paper~\cite{moed2010measuring} is a metric employed in assessing the influence of scholarly journals.  It is determined by dividing an article's citation count within a journal by the anticipated citation rate within its particular field.  SNIP considers the citation potential within the journal's discipline, enabling equitable comparisons across diverse areas of study.  In essence, SNIP serves as a valuable gauge for evaluating the impact of a journal relative to its field. It provides researchers and institutions with a standardized measure to evaluate the influence of scholarly journals rather than the impact of the specific research achievement. Furthermore, SNIP compares a journal's citation count with the citation frequency in its field. However, it fails to consider the variations in citation practices across different subject areas.  In addition, it has many inherent bibliometrics limitations we discussed in Section~\ref{bibliometrics_weakness}. 

\subsubsection{Journal Impact Factor (JIF)}
The journal impact factor, devised by Eugene Garfield, is used by Clarivate's Web of Science to evaluate a journal's impact. 
The impact factor is calculated as $C/\sum_{i=0}^{n}P_i$, where $C$ is the number of citations received in a given year for publications in a journal that were published in the $n$ preceding years, and $\sum_{i=0}^{n}P_i$ is the total number of citable items published in that journal during the $n$ preceding years. %The shortcomings of the impact factor are the unclear way in which citations are counted and the subjectivity of what constitutes a “citable item.” 

\subsubsection{SCImago Journal Rank (SJR)}
\begin{figure}
	\centering
		\includegraphics[scale=.5]{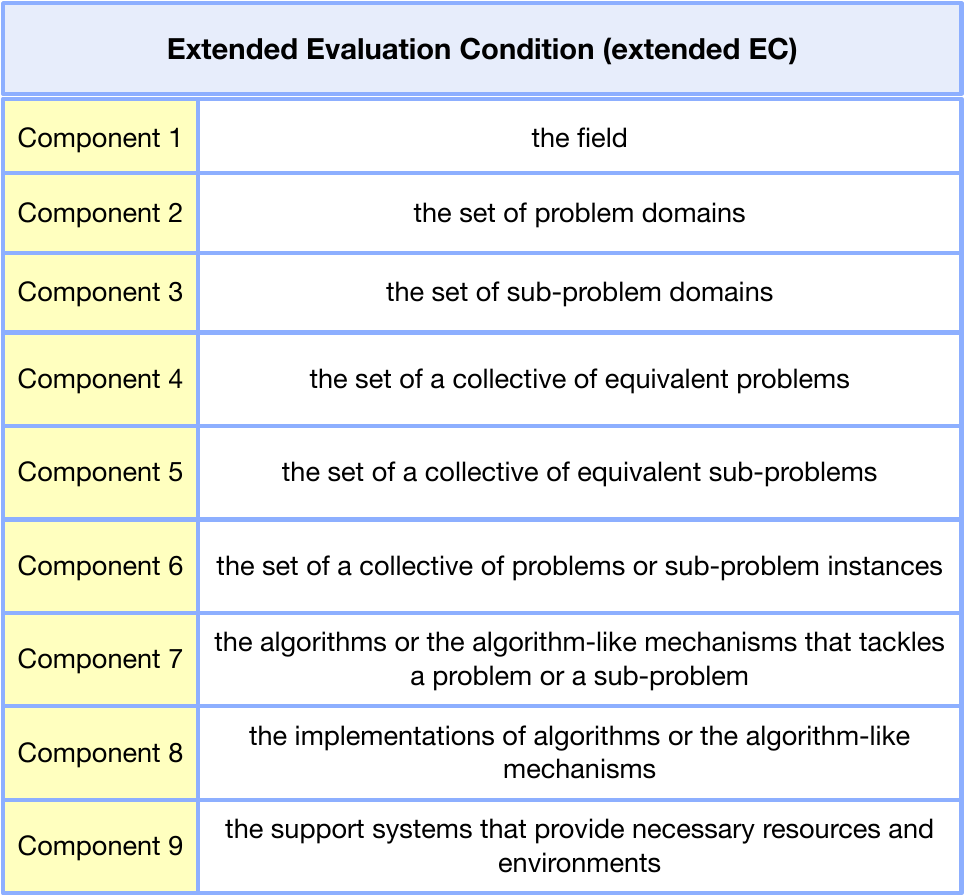}
	\caption{The overview of an extended EC.}
	\label{extendedEC}
\end{figure}
SCImago Journal Rank (SJR) indicator, developed by the Scimago Lab, is a measure of the prestige of journals.
SJR is calculated by using an algorithm similar to Google's PageRank, which assumes that important websites are linked to other important websites. Citations are used to link the journals. The algorithm begins by setting an identical amount of prestige to each journal, then using an iterative procedure to transfer each journal's achieved prestige to each other through citations until each journal's update reaches a minimum threshold. The limitations of SJR include the algorithm's complexity, the degree of transparency, and the reproducibility of the results.

Besides, Kevin W. Boyack~\cite{boyack2004mapping} utilizes data mining and analysis techniques to map knowledge domains, specifically applying them to 20 years of PNAS publications. It combines various data sources to analyze the input-output ratio and diffusion between disciplines. However, its reliance on raw citation counts as the primary measure of impact, without adjusting for self-citations, potentially leads to a skewed and less meaningful assessment of true scholarly influence.  

\subsection{The basic concepts, theories, and methodologies in evaluatology}

According to ~\cite{Evaluatology}, an individual or system being evaluated is a subject. A stakeholder is defined as an entity that holds a stake of responsibility or interest in the subject matter. 
Evaluation is "the process of inferring the impact of subjects indirectly within evaluation conditions (EC) that cater to the requirements of stakeholders, relying on objective measurements and/or testing ."~\cite{A_short_summary_Evaluatology}. 

The fundamental methodology for evaluating a single subject is outlined as follows.
 Zhan et al.~\cite{Evaluatology} propose a universal methodology to define an EC, which consists of five basic components~\cite{Evaluatology}: "(1) a set of equivalent definitions of problems; (2) the set of a collective of equivalent problem instances; (3) the algorithms or algorithm-like mechanisms; (4) the implementations of algorithms or algorithm-like mechanisms; (5) support systems that provide necessary resources and environments~\cite{A_short_summary_Evaluatology}". 

Subsequently, it becomes crucial to implement a well-defined EC for a precisely defined subject, forming a well-defined evaluation model (EM) or system (ES).

In terms of complex scenarios, the evaluation methodology is to establish a series of EMs that ensure transitivity from a real-world ES to a perfect EM and a pragmatic EM~\cite{Evaluatology}. 

Zhan et al.~\cite{Evaluatology, A_short_summary_Evaluatology} characterize the real-world ES, perfect or pragmatic EMs. Because our S\&T evaluation methodology is based on those concepts, we give a concise summary based on~\cite{Evaluatology, A_short_summary_Evaluatology}. 

The real-world ES refers to "the entire population of real-world systems that are used to evaluate specific subjects." The real-world ES has several significant obstacles: "the presence of numerous confounding,  prohibitive evaluation costs resulting from the huge state spaces."

A perfect EM replicates the real-world ES with utmost fidelity: "It eliminates irrelevant problems and has the capability to thoroughly explore and comprehend the entire spectrum of possibilities of an EC." 
However, it also has serious limitations: "possesses huge state space, entails a vast number of independent variables, and hence results in prohibitive evaluation costs."

Providing a means to estimate the parameters of the real-world ES or a perfect EM, a pragmatic EM simplifies the perfect EM in two ways: "reduce the number of independent variables that have negligible effect and sample from the extensive state space."
\section{The science and technology evaluatology}~\label{metric}

This section presents the essence of S\&T evaluatology.

\subsection{The overview}~\label{S_T_evaluatology_summary}
Understanding the development of S\&T is highly challenging. Sometimes, practice leads the way; at other times, theory does. Some individuals pose a significant problem and offer a preliminary solution, while others provide state-of-the-practice solutions without explicitly stating the problems. The relentless efforts of scientists and engineers make the landscape of S\&T achievements intricate and dense, much like an interwoven forest, thereby making the objective evaluation of S\&T contributions extremely challenging.

% Identifying the top N achievements from millions of science and technology contributions is extraordinarily challenging.

To tackle this challenge, we have adopted the evaluatology framework developed by Zhan et al.~\cite{Evaluatology} as the theoretical foundation for our research. This framework serves as the basis for developing S\&T evaluatology. The core principles and methodologies of S\&T evaluatology are outlined as follows:

\begin{figure*}
	\centering
    \includegraphics[scale=0.37]{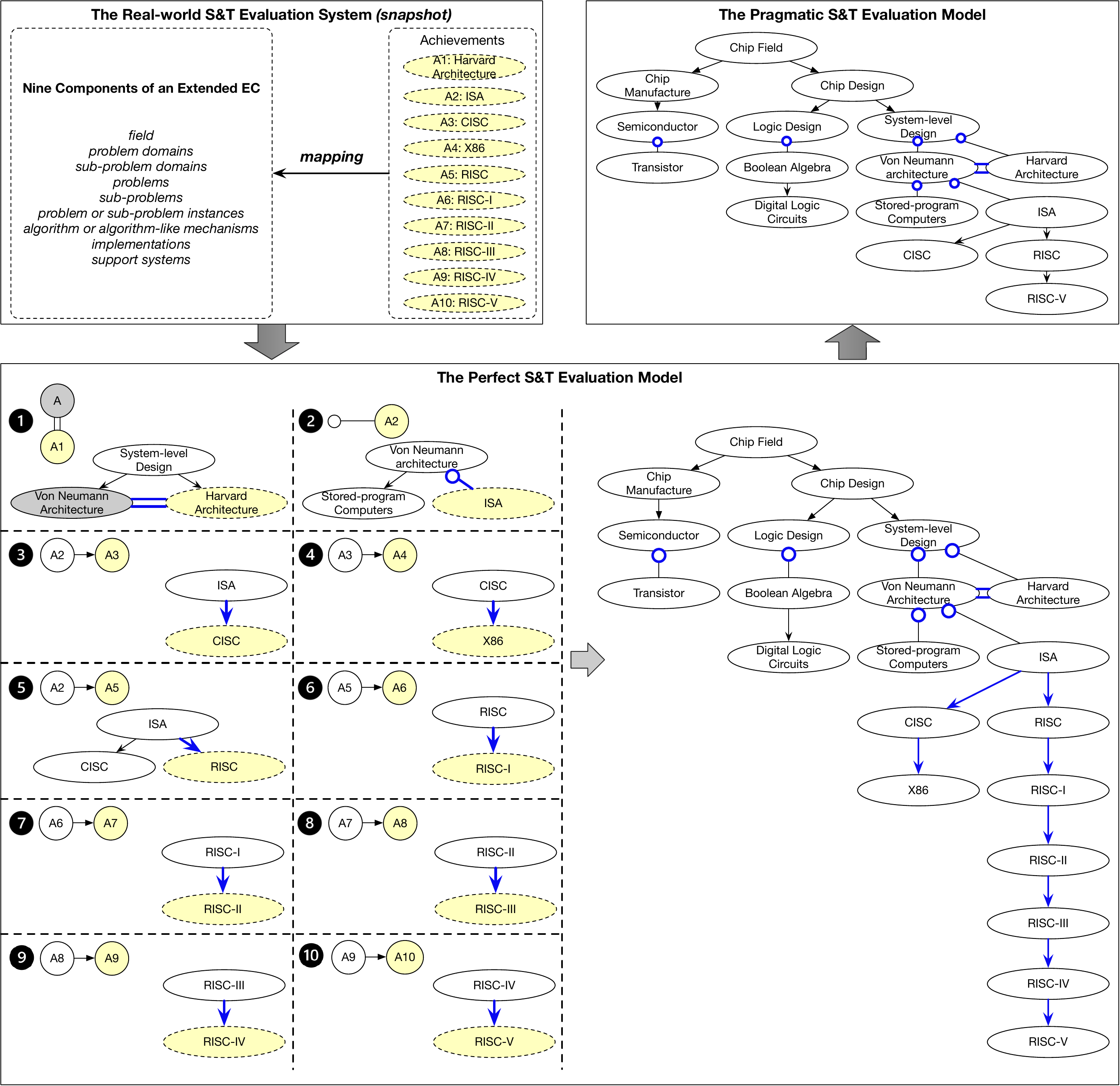}
	\caption{Illustrating S\&T Evaluatology with an example.}
	\label{st-evaluatology-example}
\end{figure*}

\begin{figure*}
	\centering
    \includegraphics[scale=0.38]{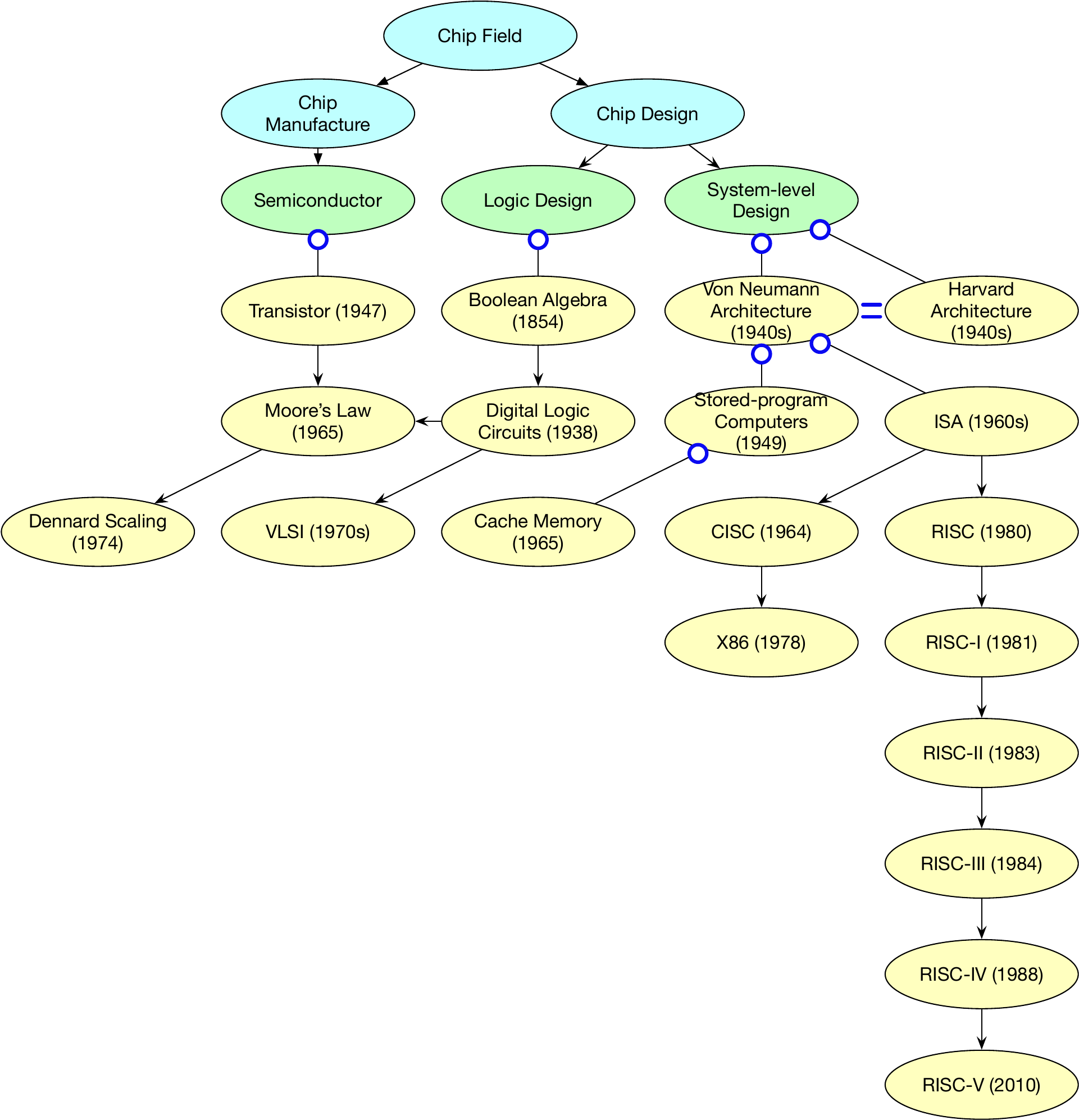}
	\caption{The localized snapshot of a pragmatic EM in the field of Chip technology.}
	\label{Chip100Overview}
\end{figure*}

First, building upon the definition of an EC proposed in the referenced paper~\cite{evaluation}, we introduce the concept of an extended EC, as shown in Figure~\ref{extendedEC}. 

With respect to the EC definition~\cite{A_short_summary_Evaluatology}, an extended EC introduces several extra components to accommodate the new requirements of S\&T evaluation, including the field that can be broken down into several problem domains, the set of problem domains, the set of sub-problem domains, and the set of a collective of equivalent sub-problems. The definition of the extended EC serves as the foundation for the proposed S\&T evaluatology. It provides the framework upon which the evaluation of S\&T achievements is based.

Second, in the realm of S\&T evaluation, a subject refers to an accomplishment that can mapped onto the nine components of an extended EC. 

For instance, let's consider a scenario where a researcher proposes a new problem and provides a preliminary algorithm for solving that problem. In this case, the subject, a specific S\&T achievement, comprises multiple components. These components include:

\begin{itemize}
    \item problem: The specific problem being addressed or investigated.
    \item Algorithm: The preliminary algorithm proposed by the researcher to solve the given problem.
\end{itemize}

Third, based on their mapped extended EC components as well as their temporal and citation links, we establish two primary relationships: ~\textit{pioneering and progressive} and two auxiliary relationships: ~\textit{parallel} and ~\textit{related but not connected} to illustrate the connections among different achievements. Section~\ref{four_relationship} will provide the details of four relationships.

Fourth, according to the theory of evaluatology, S\&T evaluation involves applying a well-defined extended EC to the subject---a specific S\&T achievement. This process allows for the creation of an EM or ES. Within a relationship under an extended EC, evaluators can effectively compare different S\&T achievements by carefully addressing the influence of confounding variables~\cite{Evaluatology, A_short_summary_Evaluatology}. 

In the subsequent four steps, we will adhere to and implement the universal evaluation methodology proposed by Zhan et al.~\cite{Evaluatology} to address the intricate S\&T evaluation scenarios. 

Fifth, we establish a real-world S\&T ES, which encompasses the complete collection of S\&T achievements. Moreover, each achievement will be decomposed into its respective components within an extended EC.   In establishing a real-world S\&T ES, it is crucial to characterize the real-world S\&T ecosystems. In line with the aim of identifying the top N S\&T achievements, the proposed real-world S\&T ES in this article encompasses the entire collection of S\&T achievements while ignoring the other components of the real-world S\&T ecosystems. 

Sixth, under the premise that all evaluated achievements belong to the same field, we assume the existence of a "perfect S\&T EM" that can accurately trace the S\&T evolution and development in terms of four relationships. That is to say, a "perfect S\&T EM" can track the evolution of a real-world S\&T ES from $ES_i$ to $ES_{i+1}$ in a rigorous manner. This model operates under the premise that only one change is made at a time. By implementing one change at a time, we ensure that only one achievement is added. 

Seventh, as the perfect S\&T EM contains huge states, we propose several simple rules to prune non-significant achievements to establish a pragmatic S\&T EM that captures the fundamental S\&T achievements. Essentially, the pragmatic S\&T EM is a collection of top N achievements. The basic idea behind this process is that we compare achievements that have a pioneering or progressive relationship under the extended EC they involve. We will explain the simple rules in Section~\ref{pragmatic_evaluation_model}.

Figure~\ref{st-evaluatology-example} illustrates S\&T Evaluatology with an example, while Figure~\ref{Chip100Overview} offers a localized snapshot of a pragmatic EM in the field of chip technology, showcasing individual achievements.
Within the chip technology field are several critical problem domains like  `Chips Manufacture' and `Chips Design'. 
This localized snapshot highlights the diversity and complexity within the chip technology field.

\subsection{The definition of an extended EC}

In~\cite{Evaluatology}, Zhan et al. emphasized that "understanding the composition of the problem domain is crucial in identifying the problem that best represents the whole. Across different disciplines, a field often exhibits a hierarchical structure, where a significant problem domain can be broken down into several problems", which provide the methodology to model an extended EC.

An extended EC consists of nine basic components~\cite{Evaluatology, A_short_summary_Evaluatology}, as shown in Figure~\ref{extendedEC}: (1) the field that can be broken down into several problem domains; (2) the set of problem domains, each of which can be broken down into various sub-problem domains; (3) the sub-problem domains, each of which can be decomposed into several problems; "(4) the set of a collective of equivalent problems, each of which can be broken down into multiple sub-problems; (5) the set of a collective of equivalent sub-problems;  (6) the set of a collective of problems or sub-problem instances; (7) the algorithms or the algorithm-like mechanisms that tackle a problem or sub-problem; (8) the implementations of algorithms or the algorithm-like mechanisms; (9) the support systems that provide necessary resources and environments~\cite{A_short_summary_Evaluatology, Evaluatology}".

\begin{figure}
	\centering
    \includegraphics[scale=0.5]{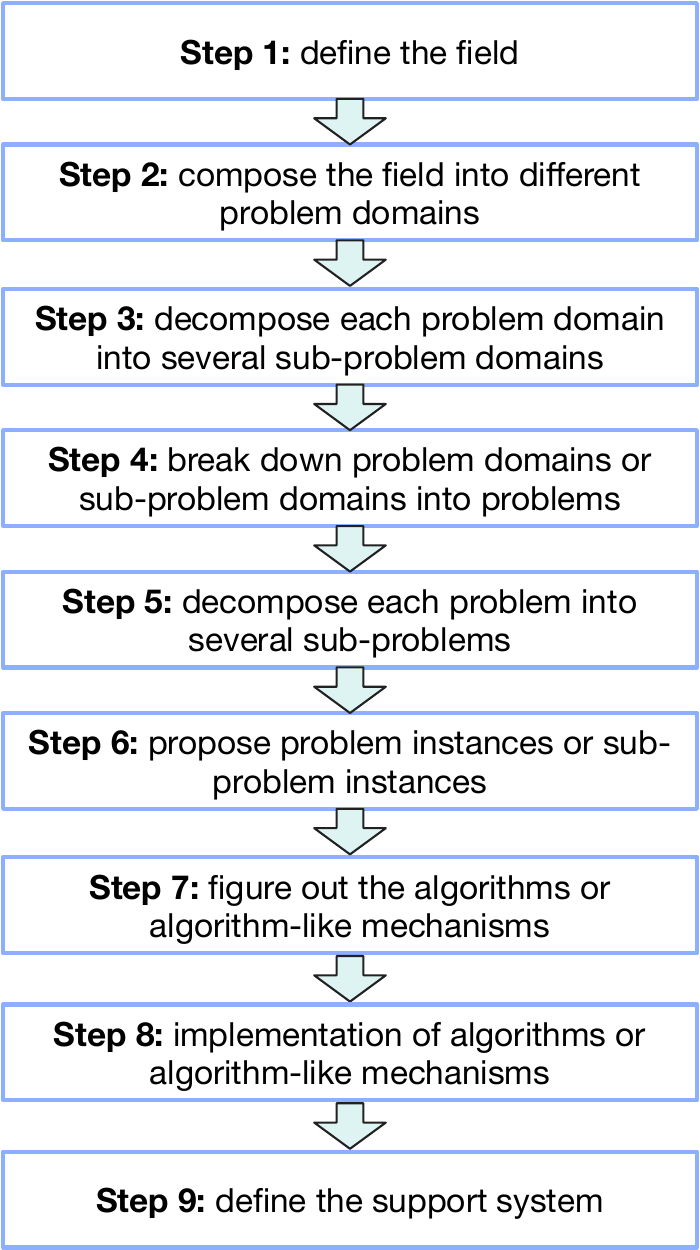}
	\caption{Essential steps of S\&T Evaluatology.}
	\label{st-evaluatology-methodology}
\end{figure}

As depicted in Figure~\ref{st-evaluatology-methodology}, the essential steps of the methodology can be summarized as follows. 
The first and second steps are to define the field and compose it into different problem domains. %($D'_r$). 
If necessary, the third step is to decompose each problem domain into several sub-problem domains. The fourth step is to break down problem domains or sub-problem domains into the problems. % ($E'_r$). 
If necessary, the fifth step is to decompose each problem into several sub-problems.  The sixth step proposes the problem instances %($E_r$) 
or sub-problem instances. The seventh step is to figure out the algorithms or algorithm-like mechanisms %($A'_r$) 
to solve the problem or sub-problem. The eighth step encompasses the implementation of algorithms or algorithm-like mechanisms. %($A_r$). 
The last step is to define the support system. % ($S_r$).}}

For example, chip design is a problem domain in the chip field. The system-level design is a typical sub-problem domain in chip design. The computer architecture design is one of the problems of the system-level design. The Von Neumann architecture was the pioneering work that defined the computer architecture design problem and proposed algorithm-like mechanisms to address it. Any specific processor that aligns with the Von Neumann architecture can be viewed as an implementation of this mechanism. 

\subsection{The formal definition of four relationships}~\label{four_relationship}

\begin{figure*}
  \centering
   \includegraphics[scale=.6]{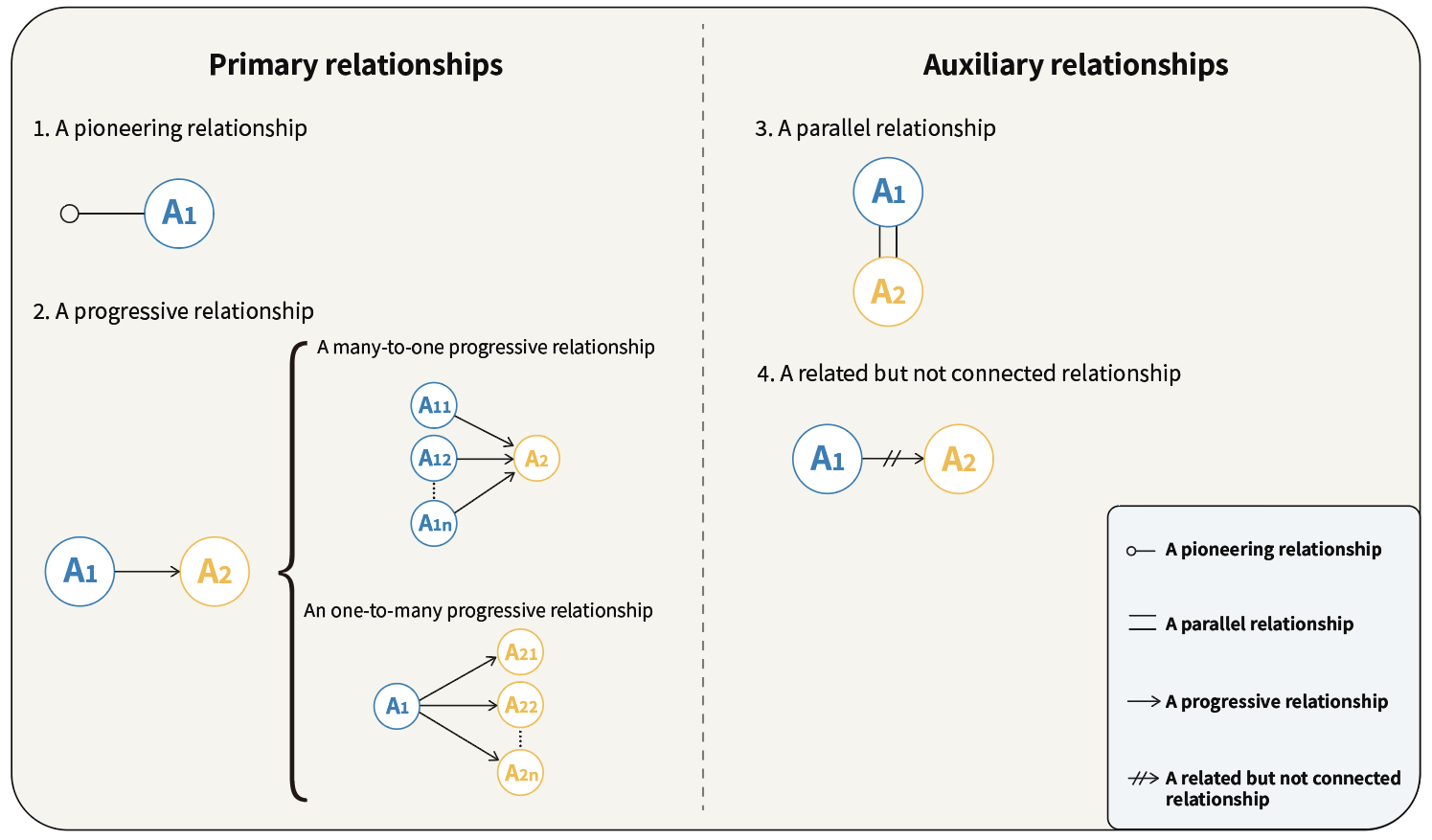}
  \caption{Two fundamental relationships and two auxiliary relationships among the S\&T achievements} 
  \label{fig:relationships}
\end{figure*}

In this section, based on their mapped extended EC components as well as their temporal and citation links, we propose two primary relationships and two auxiliary relationships to connect achievements, as shown in Figure~\ref{fig:relationships}.        

\subsubsection{Two primary relationships}

Two fundamental relationships contain a pioneering relationship and a progressive relationship.

\paragraph{Relationship One: A Pioneering Relationship.}

\textit{Definition}: A pioneering relationship pertains to an achievement that opens up a new research direction in the form of establishing a new field, problem domain, sub-problem domain,  problem, sub-problem, algorithm or algorithm-like mechanism, implementation, or support system within an extended EC.
The pioneering relationship recognizes the pioneering nature of such achievements, which lay the foundation for future advancements and innovations.

\textit{Formal expression}: Let \( A \) represent an achievement. The pioneering relationship for \( A \) can be formally expressed as:
\[
P(A) = \begin{cases} 
1 & \text{ if } A \text{ opens up a new research direction in the}\\
& \text{form of establishing a new field, problem do-}\\
& \text{main, sub-problem domain,  problem, sub-}\\
& \text{problem, algorithm or algorithm-like }\\
& \text{mechanism, implementation, or support }\\
& \text{{system within an extended EC,}}\\
%& \text{ problem, or algorithm within an extended EC.} \\
0 & \text{otherwise}
\end{cases}
%\end{array}
\]
This binary expression indicates whether \( A \) qualifies as a pioneering achievement (1) or not (0). It is based solely on the novelty and originality of the achievement \( A \), without any preceding work.

\textit{Examples}: 
Pioneering relationships manifest across various industries and disciplines, highlighting achievements that are the first to propose a novel field, problem domain, sub-problem domain, problem, sub-problem, solution, or support system within an extended EC.
\begin{itemize}
\item \textbf{Chip:} The Instruction Set Architecture (ISA) represents the pioneering work that defined the instruction set design sub-problem within the computer architecture design problem and proposed corresponding mechanisms to address it. The Reduced Instruction Set Computer (RISC) and Complex Instruction Set Computers (CISC) are subsequent developments following ISA.

\item \textbf{AI:} The first computational model of a neuron,  the McCulloh-Pitts neuron~\cite{mcculloch1943logical}, is a pioneering algorithm-like mechanism in the field of neural networks.
\end{itemize}

\paragraph{Relationship Two: A Progressive Relationship.}

\textit{Definition}: For the achievements that involve the same component of an extended EC, e.g., a problem or sub-problem, a progressive relationship indicates subsequent achievements are inspired by preceding ones, and the latter publicly acknowledges this influence through citations.

\textit{Formal expression}: A progressive relationship between two achievements \( A_i \) and \( A_j \) is defined as:

\begin{equation}
%\begin{gather}
\begin{aligned}
S(A_i, A_j) = 1 \iff \left( Q(A_i) = Q(A_j) \right) \\
\land \left( EC(A_i) \cap EC(A_j) \neq \emptyset \right) \\
%\land \left( (T(A_i) < T(A_j)) \lor (T(A_i) > T(A_j)) \right) \\
\land \left( (T(A_i\_e) < T(A_j\_b)) \lor (T(A_i\_b) > T(A_j\_e)) \right) \\
%\land \left( M(A_i) \cap M(A_j) \neq \emptyset \right) \\
\land \left( (A_i \in R(A_j)) \lor (A_j \in R(A_i)) \right)
%\end{gather}
\end{aligned}
\end{equation}

Where:
\begin{itemize}
  %\item \( T(A_i) < T(A_j) \) indicates that achievement \( A_i \) precedes achievement \( A_j \) in time.
  \item \( Q(a) \) is the key problem domain, sub-problem domain, problem, or sub-problem that achievement \( a \) addresses.
  \item \( EC(a) \) denotes the EC involved in achievement \( a \).
  \item \( T(a\_b) \) and \( T(a\_e) \) represent the begin time and end time of achievement \( a \), respectively. Thus, \( T(A_i\_e) < T(A_j\_b) \) indicates that achievement \( A_i \) precedes achievement \( A_j \) in time. \(T(A_i\_b) > T(A_j\_e) \) indicates that achievement \( A_j \) precedes achievement \( A_i \) in time. 
  \item R(a) indicates the references of achievement \( a \). Thus,  \( A_i \in R(A_j) \) indicates achievement  \( A_j \) publicly acknowledge the influence of achievement  \( A_i \).
\end{itemize}

\paragraph{A many-to-one progressive relationship}
~\\
\indent \textit{Definition}: A many-to-one progressive relationship is an instance of a progressive relationship, indicating multiple much preceding achievements inspire a single subsequent achievement.

\textit{Formal expression}: A many-to-one progressive relationship between achievements $A_{i1}$, $A_{i2}$, \ldots, $A_{in}$ and $A_j$ is defined as:
\begin{gather}
S(A_{i1}, A_j) \land S(A_{i2}, A_j) \land \cdots \land S(A_{in}, A_j) = 1
\end{gather}

Where:
\begin{itemize}
\item $\{A_{i1}, A_{i2}, \ldots, A_{in}\}$ are multiple preceding achievements.
\item $A_j$ is a single subsequent achievement.
\end{itemize}

\paragraph{An one-to-many progressive relationship}
~\\
\indent \textit{Definition}: A one-to-many progressive relationship is an instance of a progressive relationship, indicating a single preceding achievement inspires multiple subsequent achievements.

\textit{Formal expression}: A one-to-many progressive relationship between achievement $A_i$ and $A_{j1}$, $A_{j2}$, \ldots, $A_{jn}$ is defined as:

\begin{gather}
S(A_i, A_{j1}) \land S(A_i, A_{j2}) \land \cdots \land S(A_i, A_{jn}) = 1
\end{gather}

Where:
\begin{itemize}
\item $A_i$ is a single preceding achievement.
\item $\{A_{j1}, A_{j2}, \ldots, A_{jn}\}$ are multiple subsequent achievements.
\end{itemize}

\textit{Examples}: Progressive relationships demonstrate how knowledge and technology evolve over time, with each new development building on the previous ones. 
\begin{itemize}
\item \textbf{Chip:} The RISC-V instruction set architecture has its origins in and was developed from the original RISC design.
\item \textbf{AI:} %LeNet and MCDNN are two
LeNet~\cite{lecun1989backpropagation,lecun1998gradient} is a pioneering convolutional neural network that inspired AlexNet~\cite{krizhevsky2012imagenet}, a milestone in the field of deep learning.
\item \textbf{Open-sources systems:} OpenBLAS~\cite{2012openblas} is a progressive achievement of GotoBLAS2~\cite{goto2008anatomy}.
\item \textbf{Benchmarks:} The CH-benCHmark~\cite{cole2011mixed} exemplifies a many-to-one progressive relationship as it integrates aspects from both the TPC-C~\cite{TPCC111} and TPC-H~\cite{TPCH111} benchmarks. This benchmark is designed to evaluate a hybrid workload by combining the transactional operations characteristic of TPC-C with the complex querying features of TPC-H.
\end{itemize}

\subsubsection{Two auxiliary relationships}

Two auxiliary relationships contain a parallel relationship and a connected but not related relationship.

\paragraph{Relationship Three: A Parallel Relationship.} 

\textit{Definition}: A parallel relationship indicates that the achievements that involve the same component of an extended EC, e.g., problem or sub-problem, are proposed simultaneously within a brief and shared timeframe. 

\textit{Formal expression}: For a set of achievements \( A \) with each achievement \( A_i \in A \), a parallel relationship between two achievements \( A_i \) and \( A_j \) is defined as:

\begin{equation}
\begin{aligned}
P(A_i, A_j) = 1 \iff \left( Q(A_i) = Q(A_j) \right) \\
\land \left( EC(A_i) \cap EC(A_j) \neq \emptyset \right) \\
%\land \left( \left|T(A_i) - T(A_j) \right| \leq 1 \right) 
\land \left( [T(A_i\_b), T(A_i\_e)] \cap [T(A_j\_b), T(A_j\_e)] \neq \emptyset \right)
\end{aligned}
\end{equation}

Where:
\begin{itemize}
  \item \( Q(a) \) is the key problem domain, sub-problem domain, problem, or sub-problem that achievement \( a \) addresses.
  \item \( EC(a) \) denotes the EC involved in achievement \( a \).
  %\item \( T(a) \) represents the publication times of achievement \( a \) was accomplished. The achievements \( A_i \) and \( A_j \) are considered to be in a parallel relationship if the absolute difference between their publication times does not exceed one year.
 \item \( T(a\_b) \) and \( T(a\_e) \) represent the begin time and end time of achievement \( a \), respectively. %was accomplished. 
 The achievements \( A_i \) and \( A_j \) are considered to be in a parallel relationship if their time intervals overlap.  %the absolute difference between their publication times does not exceed one year.
\end{itemize}

\textit{Examples}: Parallel relationships occur across multiple fields where different approaches are employed simultaneously to address a common issue within a shared timeframe. 
\begin{itemize}
\item \textbf{Chip:} The Von Neumann architecture and the Harvard architecture are two 
parallel works in computer system-level design in the 1940s.
\item \textbf{AI:} BERT~\cite{devlin2018bert} and GPT~\cite{radford2018improving} are two parallel works in the research of big models.

\item \textbf{Open-sources systems:}  Ubuntu~\cite{ubuntuweb} and CentOS~\cite{centosweb} are two parallel works in open-source software. %open-source operating systems.
\item \textbf{Benchmarks:} BigDataBench~\cite{wang2014bigdatabench} and BigBench~\cite{ghazal2013bigbench} are two benchmarks specifically designed for evaluating big data systems, and they epitomize a parallel relationship as both were published within a year of each other, representing concurrent efforts in the problem domain of big data benchmarking. 
\end{itemize}

\paragraph{Relationship Four: A Related But Not Connected Relationship. }

\textit{Definition}: For the achievements that involve the same component of an extended EC, e.g., a problem or sub-problem, a related but not connected relationship suggests that these achievements are not proposed simultaneously within a brief and shared timeframe. Instead, they are related in some way, but there is no explicit public acknowledgment cited by the later achievements indicating inspiration or influence from the earlier ones.  

\textit{Formal expression}: A related but not connected relationship characterizes that two achievements are not parallel and have similar components inheriting the same high-level component of an extended EC but lack a citation. 

This relationship carries three implications. First, two achievements have similar components inheriting the same high-level component of an extended EC. Second, they are not parallel in nature, meaning they are not proposed simultaneously. Third, though the two achievements have a chronological order, the later ones did not cite the earlier ones. While we can not accurately disclose the underlying motivation, we emphasize the factual nature of these implications.

\begin{equation}
\begin{aligned}
C(A_i, A_{i+1}) = 1 \iff \left( Q(A_i) = Q(A_{i+1}) \right) \\
%\land \left( T(A_i) < T(A_{i+1}) \right) \\
%\land \left( |T(A_{i+1}) - T(A_i)| \geq 1 \right) \\
\land \left( EC(A_i) \cap EC(A_j) \neq \emptyset \right) \\
\land \left( [T(A_i\_b), T(A_i\_e)] \cap [T(A_{i+1}\_b), T(A_{i+1}\_e)] = \emptyset \right) \\
%\land \left( T(A_i\_e) < T(A_{i+1}\_b) \right) \\ 
%\land \left( \nexists s \in S \mid S(A_i, s) \land S(s, A_{i+1}) \right)
\land \left( (A_i \notin R(A_{i+1})) \right)
\end{aligned}
\end{equation}

Where:
\begin{itemize}
  \item \( Q(a) \) is the key problem domain, sub-problem domain, problem, or sub-problem that achievement \( a \) addresses.
  \item \( EC(a) \) denotes the EC involved in achievement \( a \).
  %\item \( S \) is the set of all achievements.
  \item \( T(a\_b) \) and \( T(a\_e) \) represent the begin time and end time of achievement \( a \), respectively.
  %\item \( \nexists s \in S \) denotes that there does not exist an intermediate achievement \( s \) in set \( S \) that forms a progressive relationship with both \( A_i \) and \( A_{i+1} \).
  \item R(a) indicates the references of achievement \( a \). \( A_i \notin R(A_{i+1}) \) indicates achievement  \( A_i \) doesn't in the reference list of achievement  \( A_{i+1} \).
\end{itemize}

\textit{Examples}: related but not connected relationships trace the sequence of achievements that tackle similar issues across different timeframes. Although these developments may seem interconnected, they often evolve independently.
\begin{itemize}
%\item \textbf{Chips:}
\item \textbf{AI:} 
%Inception~\cite{szegedy2015going} and ResNet~\cite{he2016deep} are two contemporary models for image classification.
Condconv~\cite{yang2019condconv} and Dynamic Convolution~\cite{chen2020dynamic} are two contemporary achievements for dynamical models with similar approaches.

%\item \textbf{Open-sources systems:}
\item \textbf{Benchmarks:} TPC-C~\cite{TPCC111} and TPC-E~\cite{tpc2010tpc}, both developed to evaluate Online Transactional Processing (OLTP) databases, exemplify a related but not connected relationship. They sequentially advance the field of database benchmarking without direct influence from one another.
\end{itemize}

\begin{algorithm*}
\caption{Identify four fundamental relationships among numerous S\&T achievements}
\begin{algorithmic}[1]
\State \textbf{Input:} $IDs, TimeStamps\_b, TimeStamps\_e, References, EC, ProblemsQ$
\State \textbf{Output:} $PioneerRelationship, ParallelRelationship, ProgressiveRelationship, Related But Not Connected Relationship$
\State Initialize $PioneerRelationship, ParallelRelationship, ProgressiveRelationship, Related But Not Connected Relationship$ to empty sets

\For{each achievement $i$}
    \If{$i$ opens up a new research direction in the form of establishing a new field, problem domain, sub-problem domain,  problem, sub-problem, algorithm or algorithm-like mechanism, implementation, or support system within an extended EC. }
        \State Add $i$ to $PioneerRelationship$
    \EndIf
\EndFor

\For{each pair of achievements $(i, j)$ where $i \neq j$}
    \If{$ProblemsQ[i] = ProblemsQ[j]$}
        %%% \If{\Call{TimeIntervalsWithinOneYear}{$TimeStamps[i], TimeStamps[j]$} AND $Approaches[i] \cap Approaches[j] = \emptyset$ AND $i \notin References[j]$ AND $j \notin References[i]$}     
        \If{\Call{TimeIntervalsExistOverlap}{$[TimeStamps\_b[i], TimeStamps\_e[i]], [TimeStamps\_b[j], TimeStamps\_e[j]]$} AND $EC[i] \cap EC[j] \neq \emptyset$}
            \State Add $(i, j)$ to $ParallelRelationship$
        \ElsIf{($TimeStamps\_e[i] \text{ precedes } TimeStamps\_b[j]$ OR $TimeStamps\_e[j] \text{ precedes } TimeStamps\_b[i]$) AND ($i \in References[j]$ OR $j \in References[i]$) AND $EC[i] \cap EC[j] \neq \emptyset$ }
            \State Add $(i, j)$ to $ProgressiveRelationship$
        \EndIf
    \EndIf
\EndFor
\For{each consecutive pair of achievements $(i, i+1)$, sorted by $TimeStamps\_e$}
    \If{$ProblemsQ[i] = ProblemsQ[i+1]$   AND $A_i \notin References[A_{i+1}]$ AND $EC[i] \cap EC[i+1] \neq \emptyset$ AND \Call{TimeIntervalsNoOverlap}{$[TimeStamps\_b[i], TimeStamps\_e[i]], [TimeStamps\_b[i+1], TimeStamps\_e[i+1]]$}}
        \State Add $(i, i+1)$ to $Related But Not Connected Relationship$
    \EndIf
\EndFor
\State \Return $PioneerRelationship, ParallelRelationship, ProgressiveRelationship, Related But Not Connected Relationship$
\end{algorithmic}
\end{algorithm*}

Fig.~\ref{fig:relationships} illustrates the interplay among S\&T achievements governed by four relationships: pioneering,  progressive, parallel, and related but not connected. 
In S\&T evaluatology, formalizing the four relationships is crucial for understanding and analyzing the interaction between various scientific achievements.

\subsubsection{The algorithm to identify the four  relationships}
In this subsection, we present an algorithm designed to discern four significant types of relationships among a myriad of science and technology achievements: pioneering, progressive, parallel, and related but not connected relationships. The algorithm operates on a set of inputs comprising achievement IDs, timestamps, references (key references), evaluation conditions (EC), and the key problem domain, sub-problem domain, problem, or sub-problem Q addressed by each achievement. Subsequently, it outputs sets of achievement pairs categorized into pioneering, progressive, parallel, or related but not connected relationships.

\textbf{Inputs:}
\begin{itemize}
\item $IDs$: A list of achievement IDs or an optional list of pairs of achievement IDs for comparison.
%\item $TimeStamps$: Timestamps indicating when the achievements occurred.

\item $TimeStamps\_b$: Timestamps indicating the beginning time of achievements.
\item $TimeStamps\_e$: Timestamps indicating the end time of achievements.

\item $References$: Key references or citations between achievements.
\item $EC$: The involved EC components of each achievement. %A comprehensive list of approaches for each achievement, covering motivation, designs, implementation, and experiments.
\item $ProblemsQ$: A compilation of key problem domain, sub-problem domain, problem, or sub-problem Q addressed by each achievement.
\end{itemize}

The algorithm proceeds as follows:

\textbf{1. Identification of Pioneering Relationship:}
\begin{itemize}
\item achievements that are the first to open up a new research direction by establishing a new field, problem domain, sub-problem domain, problem, sub-problem, algorithm or algorithm-like mechanism, implementation, or support system within an extended EC.
%\item achievements that have no preceding work.
\end{itemize}

\textbf{2. Identification of Parallel Relationship:}
\begin{itemize}
\item achievements addressing the same problem domain, sub-problem domain, problem, or sub-problem are scrutinized.
\item achievements occurring within 
 overlapping time intervals are classified as having a parallel relationship.
\end{itemize}

\textbf{3. Identification of Progressive Relationship:}
\begin{itemize}
\item achievements sharing the same problem domain, sub-problem domain, problem, or sub-problem are paired.
\item successive temporal order and mutual referencing between achievements, indicate a progressive relationship.
\end{itemize}

\textbf{4. Identification of related but not connected Relationship:}
\begin{itemize}
\item achievements within no-overlapping time intervals are evaluated.
\item achievements addressing the same problem domain, sub-problem domain, problem, or sub-problem, without any mutual referencing, are considered to have a related but not connected relationship.
\end{itemize}

Outputs:
\begin{itemize}
    \item $PioneerRelationship$: A set of achievement pairs in a Pioneering relationship.
    \item $ParallelRelationship$: A set of achievement pairs in a Parallel relationship.
    \item $ProgressiveRelationship$: A set of achievement pairs in a Progressive relationship.
    \item $Related But Not Connected Relationship$: A set of achievement pairs in a related but not connected relationship.
\end{itemize}

This algorithm offers a systematic approach to unraveling the intricate interplay among S\&T achievements, facilitating a deeper understanding of their underlying relationships.

\subsection{Establishing the real-world S\&T ES}

This subsection presents how to model the real-world S\&T ES ($M_r$), as depicted in Fig.~\ref{st-evaluatology-methodology}. The proposed real-world S\&T ES encompasses the entire collection of S\&T achievements, each of which is mapped onto the several components of an extended EC. As the aim is to single out the top achievements,  we ignore the other components of the S\&T ecosystem, e.g., the mechanisms and policies within the S\&T ecosystems that profoundly shape the evolution of these achievements.

Although this approach can identify all S\&T achievements, the real-world S\&T ES ($M_r$) is often susceptible to confounding factors. For instance, the communities tend to favor highly prestigious scientists, naturally drawing more attention to the research outcomes of well-known scientists. This bias stems from a real-world S\&T ES ($M_r$) 's inability to track the developmental trajectory of S\&T achievements and elucidate the relationships among these achievements. 

To address these deficiencies in the real-world S\&T ES ($M_r$), we will develop the perfect S\&T EM ($M_p$) in Section~\ref{perfect_evaluation_model}, which systematically traces the evolution of S\&T  achievements and clarifies the interconnections among them.

\subsection{Establishing the perfect S\&T EM}~\label{perfect_evaluation_model}
 
The core objective of the perfect S\&T EM is to track the evolution of S\&T achievements. This model aims to capture these achievements' dynamic changes and progressions in terms of four relationships as they contribute to the S\&T ecosystem. Doing so provides a full-picture understanding of the evolution of S\&T within the real-world context. 

Section~\ref{S_T_evaluatology_summary} has offered a concise overview of the process for establishing a perfect S\&T EM. This subsection will delve into the details, comprehensively exploring the methodology.

The perfect S\&T EM aims to track the evolution of the real-world S\&T ES. 
A perfect S\&T EM meticulously tracks the progression of a real-world S\&T ES, from $ES_i$ to $ES_{i+1}$, in a rigorous manner. This process ensures that only one achievement is added from $ES_i$ to $ES_{i+1}$.
This framework allows for an accurate description of the evolution of a field, starting from $ES_0$ and ultimately culminating in the development of a comprehensive real-world S\&T ES.

In this framework, we also provide an auxiliary structure to depict the interconnected relationships among all the achievements. As we progress from $ES_0$ to $ES_1$, from $ES_i$, then to $ES_{i+1}$, and ultimately towards a real-world S\&T ES, we adhere to the principle of adding only one achievement at a time. When a new achievement is introduced in $ES_{i+1}$, we compare it to its counterpart in $ES_{i}$ and determine the relationship based on the rules defined in Section~\ref{four_relationship}. This approach ensures a systematic and logical evaluation of the evolving achievements within the S\&T evaluation framework.

Meanwhile, as discussed in~\cite{Evaluatology, A_short_summary_Evaluatology}, the perfect S\&T EM also implies exploring and understanding the entire spectrum of possibilities within a research field. 

By embracing the concept of a perfect S\&T EM, researchers can push the boundaries of knowledge and innovation. It encourages them to explore new avenues, challenge existing assumptions, and uncover hidden potentials. Figure~\ref{st-evaluatology-methodology} shows a sample of a perfect S\&T EM. The perfect S\&T EM has almost entirely replicated the real-world S\&T ES. Not only can it establish an extended EC, but it can also organize a roadmap of achievements' evolution by identifying relationships among achievements.

\subsection{Establishing the pragmatic EM}~\label{pragmatic_evaluation_model}

Building upon the perfect S\&T EM, we can establish the pragmatic evaluation model after filtering out non-significant achievements. The process of filtering out non-significant achievements is essentially the reverse of the process outlined in Section~\ref{perfect_evaluation_model}, which explains how an achievement is added from $ES_i$ to $ES_{i+1}$. In the filtering process, we employ four rounds of filtering rules. 

In the first round, our focus is to identify and filter out non-significant achievements from those that demonstrate progressive relationships. 
For the achievements that have progressive relationships,  as they involve one or several same components of an extended EC, e.g., a problem domain or a problem, we compare achievements under the shared components of the extended EC and filter out those that are not significant. 

In the second round, we will identify the achievements that exhibit parallel relationships or related but not connected relationships to the achievements preserved in the first round. Once we have compiled these achievements, we will proceed with an additional filtering process to eliminate any non-significant ones.

We categorize an achievement that exhibits a pioneering relationship as a pioneering achievement.  In the third round, we will compare the pioneering achievements under the shared components of the extended EC and filter out achievements that are deemed non-significant.

In the fourth round, we will identify the achievements that exhibit parallel relationships or related but not connected relationships to the pioneering achievements preserved in the third round. Once we have compiled these achievements, we will proceed with an additional filtering process to eliminate any non-significant ones.

According to Zhan~\cite{zhan2022three}, an achievement can exert a positive change force over a counterpart by significantly enhancing the simplicity, user experience, cost-effectiveness, efficiency, or other fundamental features by several orders of magnitude. On the other hand, significant deviation from existing technology ecosystems can generate a negative change force. Additionally, when different usage patterns require users to incur significant learning costs, it can also result in a negative change force. This empirical law helps to explain why a certain achievement dominates over the other one.

In theory, it is possible to quantitatively measure two achievements under the same extended EC from different dimensions, and each dimension is defined as $X_{i}$. To summarize these dimensions, we propose a simple rule of thumb. We differentiate between positive and negative signs and sum up the positive or negative values of $lg(X_{i})$ (metrics from different dimensions), and the formula is shown in Equation.~\ref{formula-3-1}. This approach allows for a holistic assessment of the achievements, taking into account their various dimensions and providing a comprehensive understanding of their overall impact. By considering both positive and negative values, we can gain insights into the strengths and weaknesses of each achievement, enabling a more nuanced evaluation and comparison. 
    \begin{equation}
    \label{formula-3-1}
    V = \sum_{i} lg(X_{i}) 
    \end{equation} 

\section{The Top N @X @Y methodology}~\label{Top_N_X_Y}

As a typical case study, this section presents how to apply S\&T evaluatology.

We propose Top N @X @Y, aiming to recognize the top N achievements within a specific period X in a particular field Y. Here, N represents the number of top achievements, X represents a specific period, and Y represents a particular field. 

To optimize the effectiveness of evaluating science and technology, a standardized procedure has been devised as outlined below.

First, during a particular timeframe X, we create a real-world S\&T ES that encompasses all achievements. Each achievement is decomposed into various components within the specific extended EC.

Second, based on the real-world S\&T ES during a timeframe X, we construct a perfect S\&T EM that traces the evolution of S\&T achievements in the field of Y according to the four relationships.

Third, considering the total number of achievements (N), we assign different percentages that add up to 100\% to the achievements that have pioneering relationships and progressive relationships. 

Finally, following the four-round filtering process defined in Section~\ref{pragmatic_evaluation_model}, we filter out non-significant achievements to establish a pragmatic S\&T EM that comprises the top N achievements during a timeframe X in the field of Y. Please note that the final step is iterative. 

Following the aforementioned procedures, the top N achievements are obtained and can be presented in a tree form, as depicted in Figure~\ref{Chip100Overview}. Subsequently, we can proceed to rank these achievements along with their corresponding contributors and institutions. 

We propose a simple rule to score each achievement, with higher scores leading to higher rankings. Initially, each selected achievement is assigned a score of 1.0 points. However, we give an extra score to each pioneering achievement. With each groundbreaking achievement paving the way for new research directions, we aggregate the cumulative scores of progressive achievements by applying a weight, which we call a pioneering weight, to the original score of the pioneering achievement.

Once the scores for each achievement are determined, we proceed to assess the contribution shares of each author and their respective institutions. The specific criteria for assessing the main academic contributors are as follows:

\begin{enumerate}
    \item If the number of authors is three or fewer, the score is evenly distributed among all authors involved.
     \item If there are more than three authors, and their contributions are stated to be equal, the score is evenly divided among all authors.
      \item When there are more than three authors and their contributions are not stated as equal, the first author is assigned a first-author ratio, i.e., 0.3. In cases where multiple individuals share the first authorship, the first-author ratio is equally divided among them. The corresponding author (or the last author in the absence of a designated corresponding author) receives a corresponding author ratio, i.e., 0.3. Similarly, if multiple individuals share the corresponding author role, the corresponding author ratio is evenly distributed among them. The remaining ratio is equally divided among the other authors.
\end{enumerate}

As per the aforementioned rule, the score assigned to each achievement is subsequently distributed among the respective contributors based on their designated ratios. For every contributor, the corresponding institutions (which may be one or multiple) can be determined at the time of their contribution. In cases where a contributor is associated with multiple institutions, the score will be evenly divided among all the affiliated institutions.
\section{A Case Study on the Top 100 Chip Achievements}~\label{ranking}

The chip industry plays a crucial role in driving technological advancements across various sectors, encompassing a vast ecosystem involved in software, hardware, and application development to harness their capabilities. Utilizing S\&T evaluatology principles, the International Open Benchmark Council (BenchCouncil) has developed a well-defined extended EC to assess various aspects of chips comprehensively. The first level is the chip field, while the second level encompasses three problem domains: chip design, chip manufacturing, and chip packaging. At the third level, chip design involves several sub-problem domains, including system-level design, logic design, physical design, timing design, verification, and simulation. Chip manufacturing covers semiconductors, materials, and optics. Then, using the Top N @X @Y methodology, BenchCouncil has launched an ambitious initiative to systematically recognize and honor the most 100 groundbreaking and influential achievements in chip technology (Chip100)~\cite{Chip100}. 

The current version of Chip100 uses the Top N @X @Y methodology, where N stands for 100, X spans from the 1940s (the advent of the first computer) to 2023, and Y indicates the chip field and the percentages of pioneering achievements and 
progressive achievements are 40\% and 60\%, respectively. For the ranking in Chip100, the pioneering weight is set as 0.2, the first-author ratio is 0.3, and the corresponding author ratio is 0.3. 
%BenchCouncil is slated to release the Chip100 (1940s-2024) by the end of 2024. 

The major influential accomplishments in chips are encompassed within Chip100. For example, as depicted in Figure~\ref{Chip100Overview}, the Instruction Set Architecture (ISA) was first introduced by Frederick Brooks in the 1960s. It defines a crucial sub-problem of computer architecture design (problem) of the system-level design (sub-problem domain) within the chip design problem domain: the challenge of designing the instruction set and proposing effective mechanisms. This concept led to the development of Complex Instruction Set Computers (CISC) and Reduced Instruction Set Computer (RISC). Subsequently, Instruction Set Architectures such as X86 and RISC-V emerged, drawing from the principles of CISC and RISC. This examination provides valuable insights into the connections among these achievements. So, Chip100 identified and evaluated significant achievements and researchers in the chip field that could not be discerned through the application of bibliometrics.

\begin{table*}[]
\caption{Comparing Chip100 against CSRankings and Highly Cited Researchers from Elsevier}
\begin{tabular}{|p{2.2cm}|p{3.9cm}|p{4cm}|p{5.5cm}|}
\hline
Methods & Top 20 Achievements & Top 20 Contributors & Top 20 Institutions \\ \hline
Chip100~\cite{Chip100} \quad     (By the end of 2023)& Von Neumann Architecture, ISA, Stored-program computers, Cache memory, Boolean Algebra, Floating Point Unit, Formal Verification,
Out-of-Order Execution, Stream Architecture, Amdahl's Law, Verilog, FPGA,  Branch Predictor, CC-NUMA, 
ECC, EDA, Electrostatic Discharge, Harvard Architecture, Multi-Core Processors, NOC, SIMD Architecture, Single-Chip Multiprocessor, SOC, The Principle of Locality, and Virtual address translation

 &
John von Neumann,
Maurice Wilkes,
Frederick Brooks,
David A. Patterson,
Gene Amdahl,
George Boole,
Robert Tomasulo,
William Kahan,
Phil Moorby,
John L. Hennessy,
Aart de Geus,
Claude Shannon,
Jen-Hsun Huang,
John Gustafson,
Lisa Su,
Mark Hill,
Michael J. Flynn,
Michel Mardiguian,
Richard Hamming
Ross H. Freeman,
Wayne Wolf,
and William M. Johnson
&
Princeton University,
IBM,
Univ. of California - Berkeley,
University of Cambridge,
Stanford University,
AMD,
Intel,
Massachusetts Institute of Technology,
NVIDIA,
Xilinx,
University of Michigan,
Gateway Design Automation,
ARM,
Bell Labs,
Georgia Institute of Technology,
Google,
Harvard University,
Motorola,
Sandia National Laboratories,
Synopsys,
University of Paris South,
University of Pennsylvania,
and University of Washington
 \\ \hline
CSRankings~\cite{CSRankings} (By the end of 2023) & Achievements are predicated on the number of publications in top-tier conferences. &   	David T. Blaauw, Andrew B. Kahng,	
Srini Devadas,
Josep Torrellas,	
 Diana Marculescu,	
Mark Horowitz,  Alberto L. Sangiovanni Vincentelli,
Mahmut T. Kandemir,
Jason Cong, 
Yuan Xie, Moinuddin K. Qureshi, Giovanni De Micheli, Sheldon X.D. Tan, Onur Mutlu, David Z. Pan, Yiran Chen, ohsen Imani, Zhiru Zhang, Xiaoyao Liang, and Margaret Martonosi  & University of Michigan, Univ. of California - San Diego, Massachusetts Institute of Technology, Univ. of Illinois at Urbana-Champaign, Carnegie Mellon University, Stanford University, Univ. of California - Berkeley, Pennsylvania State University, Univ. of California - Los Angeles, Univ. of California - Santa Barbara, Georgia Institute of Technology, EPFL, Univ. of California - Riverside, ETH Zurich, University of Texas at Austin, Univ. of California - Irvine, Duke University, Shanghai Jiao Tong University, Cornell University, and Princeton University  \\ \hline
Highly Cited Researchers~\cite{highly-cited-researchers}  (2023) &  Achievements are highly cited papers& There are a total of 98 highly cited researchers in the field of computer science, listed in no particular order.& Chinese Academy of Sciences, Harvard University, Stanford University, National Institutes of Health,	Tsinghua University, Massachusetts Institute of Technology, University of California San Diego, University of Pennsylvania, University of Oxford, Max Planck Society, University of California San Francisco, University College London, University of Hong Kong,	Washington University, University of California Berkeley, 	Johns Hopkins University, Memorial Sloan Kettering Cancer Center, University of Cambridge, Yale University, 	University of California Los Angeles, and University of Washington Seattle (based on the summary of highly cited researchers from all research fields)\\ \hline
\end{tabular}
\label{Chip100_Comparison}
\end{table*}

We use the data of Chip100, CSRankings, and the Highly cited Researchers from Elsevier to find the top 100 achievements, contributors, and institutions in the chip field. 

%Chip100 uses the Top N @X @Y methodology, where N
%stands for 100, X spans from the 1940s (the advent of the
%first computer) to 2023, and Y indicates the chip field. %The percentages of pioneering achievements and progressive achievements are 40\% and 60\%, respectively. For the ranking in Chip100, the pioneering weight is set as 0.2. 
CSRankings uses the metric of the number of publications at the top-tier conferences for gauging the academic influence of researchers or their affiliated institutions in computer science.
The CSRankings database utilized by us extends across a timeline from 1970 to 2023, representing the most extensive timeframe available for CSRankings.  The most matched areas include Computer Architecture and Design Automation.

The Highly Cited Researchers list is the typical metric based on citations. The main criteria for inclusion are "the authorship of multiple Highly Cited Papers™ within the past decade and being ranked in the top 1\% based on citations in Web of Science™"~\cite{highly-cited-researchers}. Highly Cited Researchers™ represent a select group comprising only 0.1\% of researchers in the world. The data of Highly Cited Researchers utilized by us was released in the year 2023, and hence, the timeframe is from 2013 to 2023, representing the most extensive timeframe available for this database. The matched area is Computer Science, as it cannot be narrowed down to focus solely on the chip field.

Table~\ref{Chip100_Comparison} outlines a compilation of the Top 20 outcomes from Chip100, CSRankings, and the Highly Cited Researchers list published by Elsevier. Throughout the remainder of this section, we will focus on analyzing the top five achievements, contributors, and institutions from various rankings to identify any notable distinctions.

First, we contrast the results of Chip100 with those from CSRankings. 
From Table~\ref{Chip100_Comparison}, we can see that the results are totally different. 

According to the analysis conducted by Chip100 (1940s-2023), The top five achievements include the Von Neumann Architecture, ISA, Stored-program computers, Cache memory, and Boolean Algebra. These achievements are crucial in driving the development of chips. Conversely, the achievements in CSRankings are solely based on the volume of publications in top-tier conferences. 

Furthermore, the Top five institutions in the chip field encompass Princeton University
(recognized for advancements like Von Neumann Architecture, The Principle of Locality, and Virtual address translation), IBM (recognized for advancements like ISA, CISC, Amdahl's Law, and Dennard Scaling Law), UC Berkeley (known for achievements in Floating Point Unit design, RISC architecture, and RISC-V implementation), University of Cambridge (highlighted for innovations in Stored-program computers, Cache Memory, and Advanced RISC Machines), and Stanford University (acknowledged for progress in MIPS architecture, Superscalar processing, and Single-Chip Multiprocessor development).

In contrast, CSRankings only emphasizes the number of publications at top-tier computer science conferences. 
In the field of Computer Architecture and Design Automation, covering the period from 1970 to 2023, the top five research institutions include the University of Michigan, University of California-San Diego, Massachusetts Institute of Technology, University of Illinois at Urbana-Champaign, and Carnegie Mellon University. 

Among the top five research institutions selected by CSRankings, only the University of Michigan (No.11), Massachusetts Institute of Technology (No.8), and Carnegie Mellon University (No.24) are included within the Chip100 (1940s-2023), while the University of California-San Diego and the University of Illinois at Urbana-Champaign are not featured. 

The Top five contributors in Chip100  are John von Neumann (recognized for Von Neumann Architecture), Maurice Wilkes (known for Stored-program computers and Cache Memory mechanism), Frederick Brooks (credited with ISA), David A. Patterson (recognized for the monograph "Computer Architecture: A Quantitative Approach", RISC, and RISC-V), and Gene Amdahl (recognized for CISC and Amdahl's Law).
Contrasting with this viewpoint, the top five chip research contributors according to CSRankings by the end of 2023 are Onur Mutlu (117 contribution papers), Yuan Xie (116 contribution papers), Jason Cong (112 contribution papers), Alberto L. Sangiovanni-Vincentelli (108 contribution papers), and David Z. Pan (104 contribution papers). The noticeable disparity between these rankings is apparent, with none of the top five researchers in CSRankings being featured in the Chip100 list spanning from the 1940s to 2023.

Another well-known ranking is the Highly Cited Researchers published by Elsevier.  The achievements are constrained to highly cited papers as viewed through the lens of the Highly Cited Researchers. The top institutions listed in Table~\ref{Chip100_Comparison} are determined based on a roster of highly cited researchers from all research fields.
In 2023, a total of 7,125 researchers were recognized as Highly Cited Researchers, including 98 in the field of computer science. It is challenging to conduct precise searches for top institutions or researchers within a specific and focused field, such as Chip.

The criteria for this recognition clearly prioritize the impact of papers from a bibliometric perspective, as indicated by their citation counts. As a result, none of the top five contributors listed in the Chip100 have been encompassed in the Highly Cited Researchers list. On the other hand, none of the 98 Highly Cited Researchers in the field of computer science have been included in Chip100  as well.
\section{Conclusion}~\label{conclusion}
This article systematically reveals three severe bibliometrics limitations in recognizing top science and technology achievements and researchers. 
To address these shortcomings, we introduce science and technology evaluatology, which exemplifies the application of evaluatology in evaluating science and technology achievements. 
%The fundamental principle is implementing well-defined extended evaluation conditions (EC) on particular achievements to establish evaluation models or systems.
At the heart of this approach lies the concept of an extended evaluation condition, encompassing nine
crucial components. We define four relationships that illustrate the connections among various achievements based on their mapped extended EC components, as well as their
temporal and citation links: pioneering, progressive, parallel, and related but not connected. Within
a pioneering or progressive relationship under an extended evaluation condition, evaluators can effectively compare these
achievements by carefully addressing the influence of confounding variables.
The case studies show the effectiveness of the proposed methodology compared with bibliometrics.
\section{Acknowledgments}
\printcredits
% % Main text
% \input{sections/introduction}

% \input{sections/related}

% \input{sections/methodology}

% \input{sections/evaluation}

% %\input{sections/discussion}

% \input{sections/conclusion}

% \input{sections/Acknowledgments}

\bibliographystyle{elsarticle-num}
% \bibliography{cas-refs}

\begin{thebibliography}{99}

\bibitem{CSRankings}%1
Emery Berger.
\newblock CSRankings.
\newblock \url{https://csrankings.org/#/index?all&us}, 2023.

\bibitem{zhan2022three}%1
Jianfeng Zhan.
\newblock Three laws of technology rise or fall.
\newblock \emph{BenchCouncil Transactions on Benchmarks, Standards and Evaluations}, 2(1):100034, 2022.

\bibitem{hindexwiki}
H-index.
\newblock \url{https://en.wikipedia.org/wiki/H-index}, 2023.

\bibitem{highly-cited-researchers}
Highly Cited Researchers.
\newblock \url{https://clarivate.com/highly-cited-researchers/}, 2023.

\bibitem{DBLP}%1
University of Trier.
\newblock Digital Bibliography \& Library Project.
\newblock \url{https://dblp.org/}, 2023.

\bibitem{moed2010measuring}
Henk F. Moed.
\newblock Measuring contextual citation impact of scientific journals.
\newblock \emph{Journal of Informetrics}, 4(3):265--277, 2010.

\bibitem{hirsch2005index}
Jorge E. Hirsch.
\newblock An index to quantify an individual's scientific research output.
\newblock \emph{Proceedings of the National Academy of Sciences}, 102(46):16569--16572, 2005.

\bibitem{james2018citescore}
Chris James, Lisa Colledge, Wim Meester, Norman Azoulay, and Andrew Plume.
\newblock CiteScore metrics: Creating journal metrics from the Scopus citation index.
\newblock \emph{arXiv preprint arXiv:1812.06871}, 2018.

\bibitem{ioannidis2019standardized}
John PA Ioannidis, Jeroen Baas, Richard Klavans, and Kevin W Boyack.
\newblock A standardized citation metrics author database annotated for scientific field.
\newblock \emph{PLoS biology}, 17(8):e3000384, 2019.

\bibitem{boyack2004mapping}%%%%%%%kaishi
Kevin W. Boyack.
\newblock Mapping knowledge domains: Characterizing PNAS.
\newblock \emph{Proceedings of the National Academy of Sciences}, 101(Suppl\_1):5192--5199, 2004.

\bibitem{goto2008anatomy}
Kazushige Goto and Robert A van de Geijn.
\newblock Anatomy of high-performance matrix multiplication.
\newblock \emph{ACM Transactions on Mathematical Software (TOMS)}, 34(3):1--25, 2008.

\bibitem{2012openblas}
Xianyi Zhang, Qian Wang, and Yunquan Zhang.
\newblock OpenBLAS: a high performance blas library on loongson 3a cpu.
\newblock \emph{Journal of Software}, 22(Zk2):208--216, 2012.

\bibitem{evaluation}
BenchCouncil.
\newblock BenchCouncil Science and Technology Achievement Evaluation.
\newblock \url{https://www.benchcouncil.org/evaluation/}, 2023.

\bibitem{ubuntuweb}
Ubuntu.
\newblock \url{https://ubuntu.com/}, Initial release in 2004.

\bibitem{centosweb}
CentOS.
\newblock \url{https://www.centos.org/}, Initial release in 2004.

\bibitem{Chip100}
Top 100 Chips achievements.
\newblock \url{https://www.benchcouncil.org/evaluation/}, 2023.

\bibitem{cvpr2022}
Accepted Papers of CVPR 2022.
\newblock \url{https://cvpr2022.thecvf.com/accepted-papers}, 2022.

\bibitem{micro2022}
Accepted Papers of MICRO 2022.
\newblock \url{https://microarch.org/micro55/index.php}, 2022.

\bibitem{vaswani2017attention}
Ashish Vaswani, Noam Shazeer, Niki Parmar, Jakob Uszkoreit, Llion Jones, Aidan N. Gomez, {\L}ukasz Kaiser, and Illia Polosukhin.
\newblock Attention is all you need.
\newblock \emph{Advances in neural information processing systems}, 30, 2017.

\bibitem{Evaluatology}
Jianfeng Zhan, Lei Wang, Wanling Gao, Hongxiao Li, Chenxi Wang, Yunyou Huang, Yatao Li, Zhengxin Yang, Guoxin Kang, Chunjie Luo, and others.
\newblock Evaluatology: The science and engineering of evaluation.
\newblock \emph{BenchCouncil Transactions on Benchmarks, Standards and Evaluations}, 4(1):100162, 2024.

\bibitem{A_short_summary_Evaluatology}
Jianfeng Zhan.
\newblock A Short Summary of Evaluatology.
\newblock \emph{BenchCouncil Transactions on Benchmarks, Standards, and Evaluation}, 4(2):, 2024.

\bibitem{wang2014bigdatabench}
Lei Wang, Jianfeng Zhan, Chunjie Luo, Yuqing Zhu, Qiang Yang, Yongqiang He, Wanling Gao, Zhen Jia, Yingjie Shi, Shujie Zhang, and others.
\newblock Bigdatabench: A big data benchmark suite from internet services.
\newblock \emph{2014 IEEE 20th International Symposium on High Performance Computer Architecture (HPCA)}, pages 488--499, 2014.

\bibitem{ghazal2013bigbench}
Ahmad Ghazal, Tilmann Rabl, Minqing Hu, Francois Raab, Meikel Poess, Alain Crolotte, and Hans-Arno Jacobsen.
\newblock Bigbench: Towards an industry standard benchmark for big data analytics.
\newblock \emph{Proceedings of the 2013 ACM SIGMOD international conference on Management of data}, pages 1197--1208, 2013.

\bibitem{cole2011mixed}
Richard Cole, Florian Funke, Leo Giakoumakis, Wey Guy, Alfons Kemper, Stefan Krompass, Harumi Kuno, Raghunath Nambiar, Thomas Neumann, Meikel Poess, and others.
\newblock The mixed workload CH-benCHmark.
\newblock \emph{Proceedings of the Fourth International Workshop on Testing Database Systems}, pages 1--6, 2011.

\bibitem{TPCC111}
TPC-C Benchmark.
\newblock \url{http://www.tpc.org/tpcc/}, 2010.

\bibitem{TPCH111}
TPC-H Benchmark.
\newblock \url{http://www.tpc.org/tpch/}, 2010.

\bibitem{tpc2010tpc}
Transaction Processing Performance Council.
\newblock Tpc benchmark™ e.
\newblock Citeseer, 2010.

\bibitem{mcculloch1943logical}
Warren S. McCulloch and Walter Pitts.
\newblock A logical calculus of the ideas immanent in nervous activity.
\newblock \emph{The Bulletin of Mathematical Biophysics}, 5:115--133, 1943.

\bibitem{devlin2018bert}
Jacob Devlin, Ming-Wei Chang, Kenton Lee, and Kristina Toutanova.
\newblock Bert: Pre-training of deep bidirectional transformers for language understanding.
\newblock \emph{arXiv preprint arXiv:1810.04805}, 2018.

\bibitem{radford2018improving}
Alec Radford, Karthik Narasimhan, Tim Salimans, and Ilya Sutskever.
\newblock Improving language understanding by generative pre-training.
\newblock OpenAI, 2018.

\bibitem{lecun1989backpropagation}
Yann LeCun, Bernhard Boser, John S. Denker, Donnie Henderson, Richard E. Howard, Wayne Hubbard, and Lawrence D. Jackel.
\newblock Backpropagation applied to handwritten zip code recognition.
\newblock \emph{Neural Computation}, 1(4):541--551, 1989.

\bibitem{lecun1998gradient}
Yann LeCun, L{\'e}on Bottou, Yoshua Bengio, and Patrick Haffner.
\newblock Gradient-based learning applied to document recognition.
\newblock \emph{Proceedings of the IEEE}, 86(11):2278--2324, 1998.

\bibitem{krizhevsky2012imagenet}
Alex Krizhevsky, Ilya Sutskever, and Geoffrey E. Hinton.
\newblock Imagenet classification with deep convolutional neural networks.
\newblock \emph{Advances in Neural Information Processing Systems}, 25, 2012.

\bibitem{yang2019condconv}
Brandon Yang, Gabriel Bender, Quoc V Le, and Jiquan Ngiam.
\newblock Condconv: Conditionally parameterized convolutions for efficient inference.
\newblock \emph{Advances in Neural Information Processing Systems}, 32, 2019.

\bibitem{chen2020dynamic}
Yinpeng Chen, Xiyang Dai, Mengchen Liu, Dongdong Chen, Lu Yuan, and Zicheng Liu.
\newblock Dynamic convolution: Attention over convolution kernels.
\newblock \emph{Proceedings of the IEEE/CVF Conference on Computer Vision and Pattern Recognition}, pages 11030--11039, 2020.

\bibitem{zhao2023survey}
Wayne Xin Zhao, Kun Zhou, Junyi Li, Tianyi Tang, Xiaolei Wang, Yupeng Hou, Yingqian Min, Beichen Zhang, Junjie Zhang, Zican Dong, and others.
\newblock A survey of large language models.
\newblock \emph{arXiv preprint arXiv:2303.18223}, 2023.

\end{thebibliography}

\end{document}